\documentstyle[11pt,psfig,aaspp4]{article}
\newcommand{\ie}{{\it i.e.\ }}
\newcommand{\eg}{{\it e.g.\ }}
\newcommand{\degr}{\mbox{$^\circ$}}
\newcommand{\ergscm}{\mbox{\,erg\,s$^{-1}$cm$^{-2}$}}
\newcommand{\ergcms}{\mbox{\,erg\,cm$^3$~s$^{-1}$}}
\newcommand{\ergsHzsr}{\mbox{\,erg\,s$^{-1}$Hz$^{-1}$sr$^{-1}$}}

\newcommand{\ergs}{\mbox{\,erg\,s$^{-1}$}}
\newcommand{\kms}{\mbox{\,km\,s$^{-1}$}}

\newcommand{\cm}{\rm\,cm}

\newcommand{\Ho}{\mbox{H$_{\circ}$}}
\newcommand{\qo}{\mbox{q$_{\circ}$}}

\newcommand{\kmsMpc}{{\rm\,km\,s$^{-1}\,$Mpc$^{-1}$}}

\newcommand{\kpc}{{\rm\,kpc}}

\newcommand{\Msun}{\mbox{M$_{\odot}$}}

\newcommand{\OII}{\mbox{[O\,{\sc ii}]$\lambda$3727}}
\newcommand{\Oii}{\mbox{[O\,{\sc ii}]}}

\newcommand{\OIII}{\mbox{[O\,{\sc iii}]$\lambda$5007}}
\newcommand{\Lya}{\mbox{Ly$\alpha$}}

\newcommand{\NeIII}{\mbox{[Ne\,{\sc iii}]$\lambda$3869}}
\newcommand{\NeV}{\mbox{[Ne\,{\sc iv}]$\lambda$3426}}
\newcommand{\SIII}{\mbox{[S\,{\sc iii}]$\lambda$9532}}
\newcommand{\delam}{\mbox{$\Delta\lambda_{\circ}$}}

\newcommand{\fasec}{\,\farcs}

\newcommand{\fsec}{\,\fs}
\newcommand{\C}[1]{3C\,#1}
\newcommand{\FR}[1]{FR\,II\ }

\def\simlt{\mathrel{\mathchoice {\vcenter{\offinterlineskip\halign{\hfil
$\displaystyle##$\hfil\cr<\cr\sim\cr}}}
{\vcenter{\offinterlineskip\halign{\hfil$\textstyle##$\hfil\cr<\cr\sim\cr}}}
{\vcenter{\offinterlineskip\halign{\hfil$\scriptstyle##$\hfil\cr<\cr\sim\cr}}}
{\vcenter{\offinterlineskip\halign{\hfil$\scriptscriptstyle##$\hfil\cr<\cr\sim\cr}}}}}
\def\simgt{\mathrel{\mathchoice {\vcenter{\offinterlineskip\halign{\hfil
$\displaystyle##$\hfil\cr>\cr\sim\cr}}}
{\vcenter{\offinterlineskip\halign{\hfil$\textstyle##$\hfil\cr>\cr\sim\cr}}}
{\vcenter{\offinterlineskip\halign{\hfil$\scriptstyle##$\hfil\cr>\cr\sim\cr}}}
{\vcenter{\offinterlineskip\halign{\hfil$\scriptscriptstyle##$\hfil\cr>\cr\sim\cr}}}}}
\journalid{337}{15 January 1989}
\articleid{11}{14}

\slugcomment{Accepted for Publication in the {\it Astrophysical Journal}}

\lefthead{Mark J. Neeser et al.}
\righthead{The Complex Line-Emission Regions of 3C\,34}

\begin{document}

\title{A Detailed Study of the Complex \\
       Line-Emission Regions in the Radio Galaxy 3C\,34}

\vspace*{0.3cm}

\author{Mark J. Neeser\altaffilmark{1}, Klaus Meisenheimer, and Hans Hippelein}
\affil{Max-Planck-Institut f\"ur Astronomie, K\"onigstuhl 17, 69117 Heidelberg,
Germany}

\altaffiltext{1}{e-mail: neeser@mpia-hd.mpg.de}
%		 \newline Send offprint requests to: M. Neeser}

\vspace*{0.5cm}
\begin{abstract}
We present the results of an extensive Fabry-Perot investigation of the 
extended line-emission gas surrounding powerful radio galaxies.  High spatial
($\simlt$1\fasec4) and kinematical (400\kms) resolution observations of the
\OII\ emission-line of the powerful, double radio source \C{34} (z=0.689) are
given.  We present evidence that anisotropic radiation emanating from a hidden 
AGN is responsible for photoionizing a line-emission region extending more than
120\kpc.  This is indicated by the distinctively bi-conical morphology of
\C{34}'s \Oii\ gas.
%revealed by our high spatial resolution observations.  
A number of line-emission components may also show various degrees of `shadowing'
outward from the central ionizing source.
%, indicating that the radiation field
%is radiation bounded in parts of the \Oii\ region.
%This is the first detection of a distinctive photoionization cone 
%of this size 
%in a source at such a large redshift.  
A simple photoionization model
is used to show that this interpretation is energetically viable on these length
scales.  
%With cumulative gas covering factors for some of the denser line-emission
%components reaching 1.0 and lower values for the intervening gas, we compute
The luminosity of the hidden central AGN, necessary to account for the
observed \Oii\ luminosity, is compatible with that of a typical 3CR quasar
at a similar redshift.  Although this interpretation can account for the 
excitation
and most of the unique shape of the warm gas, it is insufficient to explain the 
velocity and line-width structures observed with our Fabry-Perot.  Therefore, we 
also propose that the illuminated medium surrounding \C{34} is the result of 
gas swept up by the lateral expansion and backflow of the radio source lobes.
%A close correlation between the line-emitting gas and the edges of the
%20\cm\ radio emission, as well as the detailed velocity and line width features
%of our \OII\ observations, support this interpretation.
\end{abstract}

% The different journals have different requirements for keywords.  The
% keywords.apj file, found on aas.org in the pubs/aastex-misc directory, 
% contains a list of keywords used with the ApJ and Letters.  These are 
% usually assigned by the editor, but authors may include them in their 
% manuscripts if they wish. 


\clearpage

\psfull
%\psdraft

\section{Introduction}

The distant radio galaxy \C{34} (z=0.6897) has a remarkable extended
emission-line region, whose size (140\kpc) and brightness make it ideally
suited to an in depth morphological and kinematical investigation. 
A number of high quality radio maps of \C{34}, primarily at 6\cm\ and 20\cm, 
can be found in the literature (\eg\ \cite{Jenkins77}; \cite{Garrington91};
\cite{Neff95}; \cite{Johnson95}).  These maps show this powerful radio
galaxy to be a relatively symmetric, edge-brightened, 45\arcsec\ classical 
double.  The first definitive optical identification of \C{34} was made
by \cite{Riley80}, who detected a faint, diffuse object located between the 
two brightest radio features.  Although the detection of a strong \OII\ 
emission-line was sufficient to determine the redshift of this galaxy, the 
only other feature evident on the spectra of \cite{Spinrad82} is a very weak 
\NeIII\ line.  The r$_{\rm s}$-filter and intermediate-band imaging of 
\cite{McCarthy95} 
reveal a galaxy in a rich compact cluster environment, and an extended region 
of high surface brightness \OII\ that is aligned with the radio 
source.  This is also evident in the higher spatial resolution narrow-band
\Oii\ and I-band continuum images presented by \cite{Johnson95}.
Follow-up long-slit spectroscopy (\cite{McCarthy96}) of moderate
quality detected both \OII\ and \OIII\ emission, and allowed a rough velocity
profile to be fit to the central region of \C{34}. 

\C{34} is a spectacular example of the positional coincidence
between the optical line-emission and
the radio source axis, discovered to be a common phenomena in powerful, 
high redshift radio galaxies (\cite{McCarBreug87}; \cite{ChamMil87}).
The combination of the novelty of this `alignment effect', with the often
spectacular morphologies of extended emission-line regions, and their potential 
effects on the evolution and formation of radio galaxies, attracted lively
debate, little consensus, and a large number of possible explanations for this 
phenomenon.  
However, since the line-emission regions in high redshift radio
galaxies, by virtue of their large intrinsic luminosities and large spatial 
extents, provide one of the best methods for probing the warm gas at
early epochs and across a broad range of redshifts, the importance of 
understanding the alignment effect cannot be understated.
Since this phenomena arises in extended, highly dynamic gas, an 
investigation that combines well-resolved morphologies with kinematics would 
go far in shedding new light on the question of the origin of the line-emission 
gas, its excitation, and the cause of the alignment effect.   We therefore 
began an in depth survey of the \OII\ line-emission 
in 11 radio galaxies (0.45 $\simlt$ z $\simlt$ 1.1) using a Fabry-Perot
etalon.  This instrument produces seeing-limited maps of both the morphological
and kinematical structure of these line-emission regions.
We continue with the presentation of this data 
by describing our results for the powerful, high redshift radio galaxy \C{34}.

This paper has been organized as follows.  In the first section we present
the observations and the data reduction.  In section 3 we give a qualitative
description of the emission-line and continuum morphologies, and
the kinematics of the \Oii\ gas.
Our interpretation of this data, in terms of a model to account for 
the ionization, origin, and kinematics of the warm gas in \C{34}, is discussed 
in section 4.
This section concludes with an account of alternative models.
In order to compute physical sizes, `proper' distances, and luminosities,
a standard Friedmann-Robertson-Walker cosmology with a zero cosmological
constant, \Ho=50\kmsMpc, and \qo=0.5, have been assumed throughout this
paper.  At the redshift of \C{34} this results in a luminosity distance of 
1.44 $\times$ 10$^{28}$\cm\ and an angular scale of 7.94\kpc\,arcsec$^{-1}$.

\vspace*{-3mm}
\section{Observations}

\subsection{Observations of \C{34}}

The Fabry-Perot and continuum images of \C{34} were obtained on Sept. 1991
and Sept. 1992, respectively, at the prime focus focal reducer of the 3.5\,m 
telescope on Calar Alto, Spain.  The \OII\ imaging spectroscopy was done with 
a GEC chip having 22.5$\mu$m pixels, giving a `plate' scale of 0\fasec49/pixel.  
We observed \C{34} at 10 different wavelength settings between 6270\AA\ and 
6324\AA, separated by 6\AA\ steps.  Multiple exposures of 1200 seconds
each were made at every wavelength step, and resulted in 22 individual 
Fabry-Perot images. This allowed complete coverage of the redshifted 
\OII\ line.  An interference order 30, corresponding to an instrumental
profile width FWHM of $\Delta\lambda$=8.3\AA, was chosen to give a 
spectral resolution of 395\kms.
The resulting \Oii\ images of \C{34} ranged in seeing from 1\fasec1 to
1\fasec7 FWHM, with a median value of 1\fasec3.  To investigate the continuum 
morphology, as well as
subtract its contribution to the \Oii\ images, line-free continuum exposures
were obtained immediately redward of our \OII\ observations. These were taken 
at $\lambda$=6811\AA\ ($\Delta\lambda$=180\AA), and correspond to 
$\lambda_\circ$=4033\AA\ ($\Delta\lambda_\circ$=107\AA) in the rest frame
of the radio galaxy.  A total of five exposures of 500 seconds each were 
obtained in this continuum filter using a TEK chip with 24$\mu$m pixels and 
a resulting image scale of 0\fasec53/pixel.  

\subsection{Data Reduction}
\label{sect:data_reduct}

Since the reduction procedures unique to Fabry-Perot (FP) images are described in
detail in \cite{MH92},  we will give only a brief summary of the processing
steps applied to the data obtained for \C{34}.  The initial reduction followed
that of standard CCD data processing: bias subtraction, dark current correction,
flat fielding done using dome flat exposures obtained at the wavelength 
of each Fabry-Perot science image, and the removal of cosmic ray events.
Further reduction steps, peculiar to FP imaging and necessary to extract
the kinematical information, are briefly summarized.

\begin{itemize}

\item[(i)] Night sky emission-lines (in particular, OI at 
$\lambda$$\simeq$6300\AA) cause a
`ring' of enhanced background on each Fabry-Perot
image at wavelengths shortward of the observed wavelength.
%(see figure \ref{fig:sky_emission_ring}).  
An interpolation of the background pixels
(defined from the local background level of each frame and a user-given 
$\sigma$--clipping) resulted in a fit of the radial profile of
this emission.
%along equally spaced segments around the ring.  
This background fit was smoothed in its radial and azimuthal directions and 
subtracted from each Fabry-Perot frame. 

\item[(ii)] The alignment and relative scaling of all frames was based
on a comparison of positions, profiles and intensities of 10
field stars of intermediate brightness.
Since our subsequent spectral analysis of the \Oii\ line was done over 
the entire range of sampled wavelengths and included 22
individual frames, the degree of alignment between individual 
images needed to be better than $\sim$0\farcs05 to prevent a cumulative
positional error from becoming unacceptably large.
This level of alignment was ensured by computing frame differences, examining
the residual star images, and interactively adjusting the frame shift, rotation,
and scale, until the desired overlap accuracy was achieved.
The relative intensity scaling of
each Fabry-Perot frame, with respect to the reference image, followed by
requiring that the average continuum source in the field (\ie the 10
reference field stars) showed a flat spectrum across the wavelength
range covered by the Fabry-Perot images.  
This allowed us to adequately reduce the influence of possible intensity 
gradients, characteristic of some late-type stellar spectra, in our 
determination of the normalization constant.

\item[(iii)] In order to transform all of the Fabry-Perot and continuum
images to a common positional reference frame and a common effective point 
spread function (PSF), a two-by-two pixel rebinning was done for all exposures.
The FWHM chosen for the PSF convolution (1\fasec65) was somewhat broader than 
the FWHM of the Fabry-Perot image having the worst seeing.
The result of this rebinning was a set of images
with zero positional offsets, identical intensity normalizations,
identical circular PSFs, and a pixel scale of 0\fasec247 pixel$^{-1}$.
To obtain pure line-emission images the
scaled and rebinned continuum frame was subtracted from each Fabry-Perot image.

\item[(iv)]  These line-emission images were then stacked into a cube with 
wavelength as the third dimension.
A fit of a single Gaussian profile was made through this cube at each pixel 
of the source, when the total line flux in each pixel exceeded 
a given threshold.
The output of this spectral fitting analysis consisted of six 2-dimensional
frames containing the peak amplitude of the line-emission,
the wavelength value of this maximum, its e-folding
width, and the rms errors associated with each of these 
parameters.  From these fit parameters we were then able to compute the
the relative velocity and the FWHM of the \OII\ line-emission at
each pixel.

\item[(v)]  Standard aperture photometry techniques were used to determine 
the counts--to--flux conversions from the observed standard star, and used to 
photometrically calibrate the continuum and FP images.  Observations
of the calibration standard were only made at the central FP wavelength
($\lambda$=6294\AA) setting.  However, since the remaining Fabry-Perot images 
were scaled to this wavelength (step (ii)), a calibration of the entire \OII\
line could be made. Using a generalized non-linear least-squares 
curve these flux points were fit to a Gaussian profile by varying
the amplitude, central wavelength, and standard deviation of the Gaussian
profile until an optimum fit, indicated by a minimized $\chi^2$, was achieved.
The outermost wavelength points
were assumed to sample only the continuum emission from the source, and hence
the fit was forced to be flat at this level.
The total line-emission flux was then calculated by assigning the absolute
value of the flux density obtained for the central wavelength frame to the
amplitude of the fit, and integrating the profile across the line.
The statistical uncertainty assigned to the flux density and the profile width
were found by separately varying the amplitude and the standard deviation of
the Gaussian fit, respectively, until $\chi^2$ was increased by one for each
parameter.  
We estimate a typical formal uncertainty in our determination of the total 
\Oii\ line flux to be better than 20\%.
Due to the Galactic latitude 
of \C{34} ($b\simeq -31^\circ$), no extinction correction for Galactic reddening
was necessary (E(B--V)$~<~$0.03, \cite{Burstein82}).

\end{itemize}

\section{Results}

\subsection{Morphology}

The continuum and line-emission images of \C{34} are shown as contour
maps in figure \ref{fig:3c34_lab_comp} with the components discussed in the
text labelled.  

%\placefigure{fig:3c34_lab_comp}

\subsubsection{Emission--Line Morphology}

The \Oii\ line-emission morphology of \C{34} is spectacular in both its
extent and complexity.  
The continuum-subtracted \OII\
grayscale image of \C{34} is shown in the top panel of 
figure \ref{fig:3c34_line_cont}.
%
%\placefigure{fig:3c34_line_cont}
%
The central region is dominated 
by a double-lobed, high surface brightness feature characterized by the
two emission peaks, $A$ and $B$.  
We identify the brightest line-emission source (component $A$)
with the radio galaxy, as it is displaced by only $\simeq$0\farcs3 to the
southwest of the central continuum source (see figures \ref{fig:3c34_lab_comp}
and \ref{fig:3c34_Xsection}).  In contrast, the line-emission peak at $B$
is not associated with an underlying continuum source, 
and has an intensity cross-section that is indistinguishable from a 
stellar profile.  
%
%\placefigure{fig:3c34_Xsection}
%
A further difference between these two components
is that the line-emission falls off steeply and ends abruptly to background
flux levels toward the 
northeast of source $B$.  The outer end of component $A$ levels 
off and merges with a faint finger line-emission extending a further 2\farcs5 
beyond the southwestern end of this central \Oii\ component
(see figure \ref{fig:3c34_Xsection}).  

Almost directly to the west of component $A$ lies another \Oii\ knot, component
$a$.  This source is at the apex of a unique V-shaped structure that 
opens toward the west, away from the radio galaxy center.  
It is particularly interesting that these two fingers of emission extend 
symmetrically from
either side of the line connecting component $a$ with the central radio galaxy,
and also lie on either side of the central axis of the western radio lobe.

The eastern side of \C{34} is the most extended and contains two
further knots of line-emission, components $b$ and $c$.  Component $b$
sits close to the double-lobed central feature, but marks a sudden change in
direction by being almost perpendicular to the major axis defined by components
$A$ and $B$.
Further westward from component $b$ is the strongly asymmetrical 
knot $c$, whose contours are noticeably compressed in 
the direction toward the radio galaxy center, and stretched in the opposing
direction.  This is particularly evident in the intensity cross-section of the
eastern region shown in figure \ref{fig:3c34_Xsection}.
These steep contours also continue around to the south of component
$c$, but strongly fan out toward the west and north of this source.  On this side
the surface brightness of component $c$ gradually diminishes to the fainter flux
levels of the eastern line-emission region.

%\C{34} is remarkable in that its projected \Oii\ line-emission region 
The projected \Oii\ line-emission region of \C{34} 
extends for an uninterrupted 18\arcsec\, or 144\kpc, placing it among the
largest extended emission-line regions known.  Despite this large
overall size, the lateral extent of the line-emission is quite narrow and
never exceeds roughly 5\arcsec (41\kpc), giving \C{34} an axial ratio of 
almost 1:4.
It is interesting to note the overall twisted morphology of 
the emission-line region.  Beyond the rather well-defined position angle of 
the central sources (components $A$ and $B$), the fainter more extended 
line-emission to the east and west undergo radical changes in direction.  
Considering both the differences in surface brightness and 
orientation, the overall line-emission in \C{34} can be divided into three 
distinct regions; the {\em central} component dominated by sources $A$ and $B$, 
and the two outer {\em eastern} and {\em western} regions distinguished by knots
$b$ and $c$, and knot $a$, respectively.

To improve the objectivity in determining the position angle of our 
line-emission data we included only those pixels above a flux threshold of
3$\sigma_{\rm sky\,rms}$, and computed the flux-weighted first and second
moments of the \Oii\ image.  This gave a position angle of 103\degr\ $\pm$ 1\degr
for the extended emission-line region of \C{34}.
Comparing this to the orientation of the radio emission, as measured along the
line connecting the outermost hotspots, \C{34} is moderately well-aligned
with $\Delta$PA = 18\degr.

The pure \OII\ line-emission from \C{34}, within a circular aperture of
9\farcs6 radius, is 2.23 $\times$ 10$^{-14}$\ergscm.  
%For a luminosity
%distance of $D_L$ = 1.44 $\times$ 10$^{28}$\cm\ this translates into a
%total \OII\ luminosity of
For the cosmological parameters defined in the introduction this translates to
a total \OII\ luminosity of
$$
L_{[OII]} = 5.8 \times 10^{43} ~~ {\rm erg\,s^{-1}}.
$$
If we assume that the observed \Oii\ emission originates in a plasma with
temperatures of about 10$^4$ K, in which the most abundant oxygen ion is
O$^+$, then the gas density can be related to the \OII\ line luminosity by
\begin{equation}
\label{eqn:OIIdensity}
L_{\rm [OII]} = 2.3 \times 10^{41}~n_e^2~\zeta~f_v~V_{\rm kpc} ~~ 
{\rm erg\,s^{-1}}
\end{equation}
(\cite{Osterbrock}).  Here n$_e$ is the electron density in cm$^{-3}$, 
$\zeta$ is the oxygen abundance relative to the solar value,
f$_v$ is the volume filling factor of the line emitting gas,
and V$_{\rm kpc}$ is its volume in kpc$^3$.  Assigning a value to the degree
of `clumpiness' of the gas is the greatest uncertainty in determining
the overall density, with filling factors ranging from 3 $\times$ 10$^{-4}$,
as measured from [SII] line ratios in low redshift radio galaxies
(\cite{HeckBreug84}), to 2 $\times$ 10$^{-6}$ for line-emitting gas
associated with cooling flows (\cite{HeckBaum89}).  
%For the complex 
%line-emission morphology of \C{34} we will assume the following geometry:  
%a cylinder extending the length of components $A$ and $B$ with a diameter equal 
%to the width of component $A$, for the central region; a rectangular slab having 
%the length and width of the outer, fainter, emission-line regions, with 
%thicknesses equal to the FWHM of components $a$ and $c$, for the western and 
%eastern regions, respectively. This results in a crude estimate of the volume 
%occupied by the line emitting gas of 5.1 $\times$ 10$^4$\kpc$^3$.
%
%
%  	Central region:  r=1.9 l=6.3
%  	Western region:  l=4.0 w=3.9 h=0.9
%	Eastern region:  l=7.6 w=2.5 h=0.85
%
%  V=50790.1 kpc3, zeta=0.03, f=0.00001, L=5.8x10^43
%
Taking a volume of 5.1 $\times$ 10$^4$\kpc$^3$, a
filling factor midway between the observed extremes 
(10$^{-5}$), and a relative oxygen abundance of 0.03, gives an average 
density of the \OII\ gas of n$_e$\,$\sim$\,130 cm$^{-3}$.
%
% n = 128.9102 cm^-3
%
If we assume that the gas in \C{34} is completely ionized, 
%
%and that each electron has originated from a hydrogen atom
%
the average density and volume computed above give a total mass
of warm gas in \C{34} of M$_{\rm gas}$ $\sim$ 2 $\times$ 10$^{9}$ $\Msun$.

\subsubsection{Optical Continuum Morphology}

\C{34} is equally spectacular in the light of our deep line-free continuum
image. This galaxy appears to lie in a rich, compact cluster, with 18 possible 
companion galaxies within a radius of 200\kpc\ of the central source (see
figures \ref{fig:3c34_lab_comp} and \ref{fig:3c34_line_cont}).
With the exception of component $A$, however, we do not detect \OII\ line-emission
from any other continuum source in the field despite a Fabry-Perot velocity
coverage of --1600\kms$\le \Delta v_{\circ} \le$+1300\kms, in the rest frame of 
the central galaxy.  In fact, the many 
continuum sources surrounding the central radio galaxy appear to avoid the 
contours associated with the line-emission image.

Near to the center of the cluster lies the brightest source,
component $A$.  This galaxy is positioned almost exactly at the center 
of the radio emission and is identified as the optical counterpart to the 
radio source (\cite{Riley80}).  From the line-emission associated with
component $A$ we compute a redshift for this central source of
z=0.6900 $\pm$ 0.0007.  As is evident from figure \ref{fig:3c34_Xsection}, 
the core of component $A$ is only marginally different from a stellar profile.  
However, beyond a radius of about 2\arcsec\ $A$ appears to be embedded in a 
broad continuum halo, within which a number of faint knots of emission can be 
seen.
The central continuum component $A$, as derived from our continuum
frame and 17 reference stars from the PPM catalogue (\cite{Roser91}),
has an absolute position of
$$
\alpha = 01^{\rm h}~07^{\rm m}~32\fsec55 ~~~~~~~ 
\delta = +31\degr ~31^\prime ~22\fasec 8 ~~~~~(1950.0),
$$
with an uncertainty of $\pm$0\fasec5 in both coordinates.
This position was used to place the radio contours in figure 
\ref{fig:3c34_line_cont} and agrees to within 0\fasec6 of the radio nucleus
detected in the $\lambda$6\cm\ and $\lambda$20\cm\ observations of
\cite{Johnson95}.

%4\farcs4 (34.5\kpc) 
To the north of the radio galaxy lies a close (at least 
in projection) companion galaxy, component $C$.  Although this companion is 
close enough to component $A$ to share its outer halo emission at the lower 
flux levels, it is undoubtably a distinct source.  
%
%At flux levels 5.5 times the rms fluctuation above the background, this 
%connection disappears.  
%
Inspection of the line-emission images at the faintest flux levels and at all 
wavelengths, revealed no \OII\ counterpart to this continuum source.  
Therefore, we cannot assign a redshift to this component.  
%In contrast to the 
%asymmetry observed in component $A$, primarily the result of its diffuse halo, 
%source $C$ is quite symmetrical, with only a slight north-south elongation.  
%Laterally to this direction, the cross-section of component $C$ is 
%indistinguishable from a stellar profile.

In contrast to the line-emission image, the crowded continuum environment of 
\C{34} makes an unequivocal determination of the position angle of the host
galaxy difficult.  
Beyond a radius of $\sim$3\arcsec\ from the central radio galaxy the position
angle changes sharply,
as a number of companion sources, in particular component $C$, enters the
aperture.  
However, since the companion source $C$ is distinct from the radio 
galaxy at the brighter flux levels with no associated \OII\ detection,
we determined a position angle that is confined to the direct environment of 
component $A$.
Within an aperture of 3\farcs0 radius centered on this source, with component
$C$ subtracted from the image, the continuum position angle was found to be 
106\degr\ $\pm$ 5\degr.  Thus the continuum emission is well aligned
with the line-emission ($\Delta$PA($|$line--continuum$|$)=3\degr) and only
moderately aligned with the radio source 
($\Delta$PA($|$radio--continuum$|$)=21\degr).

In table 1 we present a
synopsis of the relative positions, physical separations, continuum magnitudes, 
\OII\ flux levels, and FWHM of the continuum and line-emission components 
observed in \C{34}.
%
%\placetable{tab:3c34_components}
%\centerline{EDITOR:  PLACE TABLE 1 HERE}

\subsection{Kinematics: Radial Velocities and Deconvolved Line Widths}

Figure \ref{fig:3c34profiles} illustrates line profiles at specific
positions in the line-emission region of \C{34}.
The velocity map obtained from the fit to the Fabry-Perot data cube
is shown in figure \ref{fig:3c34velwidt}, with the velocity and FWHM profiles
along the three major regions of \C{34} shown in figure
\ref{fig:3c34velfwhmpro}.  The kinematical and positional origins of 
figures \ref{fig:3c34velwidt} and \ref{fig:3c34velfwhmpro} are defined by the 
radio galaxy center.
%
%\placefigure{fig:3c34profiles}
%\placefigure{fig:3c34velwidt}
%\placefigure{fig:3c34velfwhmpro}
%
%
Compared to the other sources in our overall sample the range of velocity 
spanned by \C{34}, in particular considering the extreme size
of its emission-line region, is quite moderate.
The dynamical structures visible in this source, based on their shape and
absolute velocity, can be divided into the same three distinct regions that
were used in describing the line-emission morphology.
The velocity of the \Oii\ gas in the central region of \C{34} 
(figure \ref{fig:3c34velfwhmpro} central panel) is distinguished by a relatively
smooth velocity bump spanning 0\kms\ near the center of this region, 
to -500\kms\ at its northeast and southwestern ends.  The line-emission sources
$A$ and $B$ are positioned on either side of this feature. 
The corresponding line widths show an obvious minimum at the position of this 
velocity peak, which is $\sim$1\arcsec\ to the northeast of component $A$.
This local FWHM minimum is at 300\kms\ and
rises steadily away from both $A$ and $B$ to $\sim$750\kms\ and 550\kms\,
respectively.  The subsequent decline in the FWHM observed in
figure \ref{fig:3c34velfwhmpro}, beyond about -3\arcsec\ and +2\arcsec, may be
due to low flux levels resulting in poor fits to the data cube, or to a slight
over-subtraction of the continuum.  

The eastern region of \C{34} is surprisingly flat at velocity values near
--400\kms.  The fact that the deviations from this velocity value are always 
less than about 150\kms\ is remarkable considering that this side of the source
spans more than 70\kpc\ of projected line-emission.  The associated 
velocity widths of this region reveal a number of low significance
undulations.  The FWHM on this side of the \C{34} are the lowest
for the entire source and, in an absolute sense, the average FWHM ($\simlt$ 
250\kms) of this region is remarkably quiescent.

To the west of \C{34} we observe a very sudden velocity change in going from
the gradual decline still visible from the central region, out to component $a$.
This velocity discontinuity spans about 690\kms\ from the western end of
the central region to the beginning of source $a$.  This is most evident in
our two-dimensional velocity images (figure \ref{fig:3c34velwidt}) and points
to the fact that we are seeing an overlap of two distinct sources (\ie\
component $A$ and component $a$).  This superposition of sources results in 
the large FWHM seen at this position.  At the location of component $a$, we see 
the maximum overall velocity for \C{34} (v=500\kms). 
Across the remainder of the western region the velocity remains relatively flat.

\section{Discussion}

\subsection{Photoionization by a Hidden AGN}

Most noteworthy about the morphology of \C{34} is that the line-emission
exhibits a bi-conical shape with the nucleus of this radio galaxy at the
apex of the cones.  We will argue that this
unique shape is the result of an ambient medium swept aside by the radio source,
and subsequently illuminated by a collimated source of ionizing radiation.
%
%\placefigure{fig:3c34cone}
%

Since the likeliest location of the active nucleus lies at the centroid of
continuum component $A$, we will designate this as the origin of the cone.  
We will also assume the photoionization cone to be close to the plane of the
sky; an assumption consistent with the narrow gap between the radio lobes and
the evidence for jets on both sides of the source (\cite{Johnson95}).
As is shown
in figure \ref{fig:3c34cone}, it then immediately follows that a straight line
can be drawn from the peak of component $c$, through the origin, to the peak 
of the finger of emission northwest of component $a$.  Similarly, a straight 
line can be drawn through component $B$, the origin, the line-emission 
component $A$, and the center of the faint extension of
line-emission at the southwestern end of component $A$.  Thus, simply by
placing the apex of a symmetrical bi-cone at the position of the 
central continuum source, we find that we can connect 6 distinct line-emission 
knots/extensions on both sides of \C{34} and symmetrically straddle the radio 
source axis.  Drawn in this way,
the bi-cone has an opening angle of 60\degr.  However, since it is defined
by the positions of the peak intensities of the brightest emission-line
components, the true opening angle necessary to photoionize all of the
\Oii\ gas may be somewhat larger.  A bi-cone drawn through the same center, 
but using the outer FWHM positions of components $B$ and $c$, results in an 
opening angle of about 90\degr.

Such a conical structure in \C{34}, though unobserved
in high redshift radio galaxies, has been observed in low redshift
Seyfert galaxies and in 3C\,227 (z=0.085) (\cite{Prieto93}).  Currently 11 such 
Seyferts are known to possess an
ionization cone or a bi-cone---8 of which also contain a linear radio structure 
(\ie\ a double, triple, or jet-like radio source) (\cite{Wilson94}).
From these sources \cite{Wilson94} find that there is a tight alignment
between the cone and radio axes, with a mean difference in position angle
of only 6\degr, and that the degree of collimation is much better for the
radio plasma than for the ionizing photons.  This is exactly what is observed
in \C{34}.  As in the Seyferts, the bi-cone of \C{34} is situated
almost exactly along the axis defined by the radio emission, with 
$\Delta$PA = 4\degr. 
%And, even when measuring from
%the outer contours of the 20\cm\ emission lobes, the radio source
%opening angle is less than 35\degr.
The prominent cones in Seyferts and 3C\,227, however, show emission
across the entire lateral extent of their opening angles (see, for example,
the \OIII\ image of NGC\,5252 in \cite{Tadhunter89}).  This is
not true for \C{34}, where the line-emitting gas appears to be
confined to the outer edges of the ionization cone and by no means fills it.
%A possible explanation for this difference will be given below when we
%discuss the details of the photoionization model for \C{34}.

Aside from explaining the overall conical morphology of \C{34}, the 
photoionization model can be applied to a number of smaller-scale features
seen in line-emission.  
The intensity profiles of components $c$ and $a$ on the 
eastern and western side of the bi-cone, respectively, are both markedly 
steepened on their sides facing the central source 
(see figure \ref{fig:3c34_Xsection}).  This could be due to geometrical dilution
or by absorption or scattering in intervening material closest to the central
source.
It is also possible that radiative bounding of the ionizing flux from the central
galaxy can give rise to the morphology seen in source $B$, and in the
western component.  The steep fall-off in intensity and the lack of any
line-emission outward from component $B$ points toward this source being a
very dense knot of gas capable of absorbing all of the incident ionizing
radiation.  It is also plausible that component $a$ is `shadowing'
the ionizing emission from the AGN, and preventing this flux
from reaching the gas to its immediate west.  The small lateral extent of source
$a$ may allow the ionizing radiation to reach the gas surrounding it
to the north and south, causing the two fingers of \Oii\
emission observed extending to either side of component $a$.  The ionizing
radiation reaching the gas on the western side of \C{34} could be somewhat
diluted due to strong absorption occurring in the central component $A$.  
This would, in turn, allow component $a$ to effectively absorb the remaining 
radiation and at the same time remain relatively faint.
An alternative explanation for the fall-off in \Oii\ emission 
outward from components $B$ and $a$ could, of course, be a simple paucity 
of gas beyond these sources.  

A critical test of our photoionization model is whether or not the central,
hidden AGN can provide enough ionizing photons to excite the \Oii\ emitting
gas observed in \C{34}.  The two best line-emission sources on which to test 
this are component $c$, because it is furthest from the central galaxy, and
component $B$, because it is the brightest source beyond the central
galaxy.  The very sharp fall-off in \Oii\ brightness on the side of component
$B$ facing away from the central source, argues for a region that
is radiation bounded.  Furthermore, since the
\Oii\ flux integrated across the eastern component out to source $c$ is
almost identical to that observed for source $B$, we will assume that
$c$ is also radiation bound.  Thus, a covering factor of unity will be used for 
these two sources.

A description of the model that we have used to estimate the luminosity of 
the central source, necessary to photoionize the observed line-emission regions, 
is given in the appendix.  This model assumes that the central ionizing source 
has a spectral index of 1.0 and that the ratio of ionizing 
photons to resulting \OII\ photons is 6:1.  Then, using
equation \ref{eqn:NionAGN4} we have computed the required luminosity 
of the central source necessary to produce the \OII\ flux observed in components 
$B$ and $c$.  The results of this computation are summarized in 
table 2 using source radii based on both the FWHM and 
the total lateral extent of components $B$ and $c$, and two different ionization 
cone opening angles.  Clearly, component $c$ is the most critical source due 
to its distance from the central galaxy.  Even with the worst case parameters 
of a large 88\degr\ opening angle and a small radius, we conclude that the 
required luminosity of the hidden, photoionizing AGN is reasonable, at roughly 
a factor of three lower than the B-band luminosity of the brightest 3C quasar.
%(\ie\ \C{9} with L$_B \simeq$ 6 $\times$ 10$^{45}$ \ergs).  
If viewed along a 
line-of-sight down the ionization cone the apparent V-band magnitude (see
column 6 of table 2), assuming 
zero extinction, would be comparable to a typical 3C quasar at the redshift of 
\C{34} (\cite{Allington-Smith84}).
%
%\placetable{tab:3c34photo_model}
%
%
%
% L(B) (brightest 3C qso: 3C 9) ~ 6.2 x 10^45 erg/s====NEW
%
%

Of course, the evidence for photoionization in \C{34} could be
improved using line ratio diagnostics.
For such tests a number of low-to-high ionization lines would have to be
detected in the faint, extended line-emission region (\eg\ component
$c$).  The literature shows, however, that even the detection of a sufficient
number of lines in the 
considerably brighter nucleus has been difficult; the high ionization
lines \NeIII\ and \NeV\ were found to be very weak and undetectable, 
respectively, in the spectrum of \cite{Spinrad85}, and \cite{Rawlings91} failed 
to detect the \SIII\ line in their near-infrared spectrum.
%In future, we will attempt to repeat our FP observations with \OIII\
%to discern whether this higher ionization emission-line is less extended
%than \OII, as would be predicted by a central photoionization model.

\subsection{The \Oii\ Kinematics and Radio Source Expansion/Backflow}

The final issue that must be addressed is how the velocities observed in the 
\Oii\ gas of \C{34} fit into the photoionization framework.  The simplest 
model would assume that the ionization cone is merely illuminating gas clumps,
of random velocity, that exist in a typical cluster environment.  The abrupt 
velocity discontinuities 
between the central source and the two line-emission regions to the east and
west would tend to support such a picture.  The overall velocity spanned by the
entire line-emission region 
%(roughly -400 to +500 \kms along the line-of-sight)
is also consistent with a rich cluster (\eg\ \cite{Dressler92}).
%({\it cf.} \cite{Dressler92} who find a velocity dispersion of 1300 \kms\ 
%in the z=0.46 cluster surrounding the radio galaxy \C{295}).  
The difficulties with
this interpretation, however, arise when one considers that the \Oii\
line-emission in \C{34} is uninterrupted across more than 140\kpc---unusual
considering that the warm gas in a cluster environment is usually considered
to be in discrete clumps or filaments unconnected over such large regions.
Even more fatal to the cluster interpretation is the fact that the 
eastern component shows a remarkably flat velocity structure across a length 
of more than 70\kpc.  Such uniformity across such large distances is 
incompatible with the essentially random velocity dispersions associated 
with clusters.  In fact, the constant velocity across the entire eastern
line-emission region would seem to indicate that a {\em single} mechanism is 
required to act on a large fraction of gas simultaneously.

As is clear from figure \ref{fig:3c34_line_cont} the eastern line-emission
region is near the outer edge of the radio source, and also seems to conform
well to the radio contours.  We therefore propose that the radio source has,
through the bulk motions of its lateral expansion, enmasse swept
up the gas that existed in the environment of \C{34}.  In this way the
gas that makes up the eastern \Oii\ line-emission region is compressed,
pushed to the outer edge of the radio, and given a bulk velocity 
that is constant across the entire region.  
The connection between the radio lobes and the line-emission
gas is corroborated by the orientation of \C{34}, independently deduced from
the radio observations and our optical data.   

Radio polarization measurements of 
\C{34} by \cite{Johnson95}, find a higher degree of depolarization in the 
eastern radio lobe, particularly prominent along the position of the 
line-emitting gas.
The western side of \C{34}, on the other hand, shows no similar
match between the location of its line-emission gas and depolarization in
the radio lobe.  This supports our radio lobe expansion and
backflow interpretation of the blue-- and redshift observed in the eastern 
and western line-emission regions;  namely, that the 
line-emission to the east of \C{34} is oriented toward us, while the western 
region points into the plane of the sky.

Lateral expansion in the radio source lobes may also be 
playing a significant role in defining the velocity observed in the western
line-emission region.
In this case its redshifted velocity with respect to the central galaxy could
be due to the far side of the radio lobe expanding away from us and sweeping
the gas along with it.  The fact that the western line-emission region is
symmetrical with respect to the radio axis may be due to the gas
spreading both north and south as the far side of the radio lobe sheath
expands.  
%Combining this with some amount of radio lobe 
%backflow along the main axis, but on the outer sheath of the radio lobe, 
%would result in the western line-emitting gas being pushed back near
%component $a$, at the same time that it is stretched in the lateral direction 
%because of lobe expansion.  
%The different line-of-sight velocity between
%the projected central axis of the radio cocoon and its upper edge can explain
%the lower recessional velocity of the \Oii\ finger northward of component $a$.
Alternatively, the western line-emission region, as is indicated by the
velocity and line width discontinuity between this region and the central
source, may simply be a distinct gas cloud that is being illuminated by
the ionization cone of \C{34}.  
%The velocity structure along this feature
%would then be simply that of the random environment surrounding the radio
%galaxy.  
%If we require that component $a$ be within a cone with an opening
%angle of 60\degr, in the plane of the sky, then the gas of the western
%line-emission component will be $\simlt$20\kpc\ behind the radio galaxy.
We are currently unable to distinguish between these two interpretations.

More difficult to explain is the velocity structure observed in the central
line-emission region.  The smooth velocity bump may indicate some form of
mass inflow or outflow, whose asymmetry may be intrinsic, or due
to some form of extinction that inhibits our view of the far-side velocity
surface.  Alternatively, if we assume that the radio lobes lie behind both
the northeast and southwest ends of this component, but do not extend all the 
way to the center of the source (as seems to be indicated in figure
\ref{fig:3c34_line_cont}), backflow and/or cocoon expansion could
selectively push these ends toward our line-of-sight and create the observed
velocity bump.  Radio source backflow is indicated by the
prominent `wings' of radio emission near to the central source, that
extend perpendicular to the major axis (see figure \ref{fig:3c34_line_cont}).
The close correspondence between the radio source and the
NE end of component $B$, as well as the velocity width maxima observed at the 
ends of both central components (see figure \ref{fig:3c34velwidt}), could be 
the result of such an interaction.  It is interesting to note that
the NE line width maximum is not symmetric with respect to the major axis of the
central component ($A$ + $B$), but lies along the radio source axis.

In our photoionization scenario there exists a direct
cause and effect relationship between the radio ejecta and the ionization cone 
that leads to their alignment.
It is possible to imagine that the nucleus was initially surrounded 
by a cloud opaque to ionizing radiation in all directions.
When the radio jet turned on it plowed through the cloud and opened up a
low density channel.  As the radio lobes grew in size the increased density 
of the swept up gas at its outer edges allows it to effectively absorb
the incident ionizing radiation from the central AGN which, in turn, can
effectively escape along the cleared out, low density channel created by the 
radio source.
If this is indeed the case, one might expect that the line-emitting gas
would be edge-brightened on its radio source side due to the enhanced density
along this interface.  The intensity contours of our \OII\ map show the exact
opposite; the eastern line-emission region has a higher surface brightness on 
the side facing away from the central axis of the radio source.
If we assume that the gas of the eastern line-emission region wraps 
slightly around the bottom surface of a cylindrically
shaped radio lobe (an interpretation consistent with the depolarization at
this position), then the brightening we perceive at the outer edge of
the \Oii\ emission could be due to foreshortening.  Instead of looking
at a thin sheath of gas, as along the northern side of this line-emission
region, we see a slightly greater column depth of gas at the bottom edge of this
feature (see figure \ref{fig:3c34_edge_bright}).
%
%\placefigure{fig:3c34_edge_bright}
%

The evidence for directed radiation from a hidden nucleus provides
supporting evidence for the `unified models', that attempt to explain the
relationships between active galaxies and quasars in terms of anisotropic
emission and orientation effects
(\eg\ \cite{Barthel89}; \cite{Antonucci93}).
The high redshift and very extended line-emission of \C{34} makes this source
an important example for this class of object.
In light of this, it is important to ask the question why the morphology
of the line-emission gas observed in \C{34} is so different from the
ionization cone-filling morphology common in Seyfert galaxies.  The answer
to this is twofold.  
First, we will use NGC~5252 (cz=6852\kms) as a suitable comparison; it 
represents a well defined example of photoionization in Seyferts, and has the 
largest known ionization cone with a total extent of roughly 36\kpc.
In our image of \C{34} this would correspond to the first 2\farcs3 
on either side of the source and hence be well within the line-emission
surrounding components $A$ and $B$.  At this inner position it may well be
that the line-emitting gas fills the lateral extent of the ionization cone.
Spatial resolutions better than $\sim$0\fasec4 FWHM 
would be necessary to adequately resolve the warm gas this near to the 
central galaxy. An alternative explanation, however, along the lines of the 
`unified models', would be that the ionization cones in both NGC~5252 and 
\C{34} began in similar environments with both initial radio sources clearing 
out channels in the gas that eventually allowed the photoionizing radiation to 
escape.  However, whereas the radio source in NGC~5252 has 
P$_{\rm 6cm}$ = 3.4 $\times$ 10$^{28}$ \ergsHzsr\ and is very modest in
size (\cite{Tadhunter89}), \C{34} is more than 10$^{4}$ more 
powerful at this wavelength.  It is therefore plausible that only the 
considerably more powerful radio source in \C{34} (and high redshift \FR\ 
radio galaxies in general) is capable of effectively sweeping out the IGM of 
its host source and confining the line-emission gas to its edges.  

\subsection{Alternative Interpretations}

The ability of our photoionization model, in combination with radio source
backflow and expansion, to explain the observed line-emission morphology 
and kinematics, does not guarantee that it is the only possible interpretation
of \C{34}.  We therefore feel obliged to account for possible alternative
models, and discuss why we consider them to be inferior to our model in 
explaining our observational results.

\subsubsection{Radio Source Collisional Excitation}

The fact that the line-emission gas in \C{34} extends great distances on
either side of the central radio galaxy, the average overall alignment between 
the \Oii\ emitting gas and the radio source, and the general red-to-blueshift 
velocity trend across the source, would seem to point to a direct 
collisional interaction between these two media.  A closer look at a number 
of detailed features, however, can rule out this scenario as the main
mechanism for ionizing the \Oii\ gas.

Although there does exists an overall red-to-blueshift trend in going 
from west to east in \C{34}, the velocity changes are abrupt and 
then remain almost constant across the length of the eastern line-emission 
region.  Along the entire 70\kpc\ of this \Oii\ region the total velocity 
variation is less than 150\kms and is associated with moderate line widths of 
only about 250\kms.  This gives \C{34} a velocity profile and line widths
unlike those observed in powerful radio galaxies in which jet--cloud
interactions are more obviously involved in creating the observed emission-line 
regions.  At low redshifts examples of such sources include 4C\,29.30, 3C\,171, 
and PKS\,2250--41 (\cite{vanBreugel86}; \cite{HeckBreug84}; \cite{Clark96}), while
for high redshifts 3C\,352 and 3C\,368 (\cite{MH92}; \cite{HM92}) are obvious
candidates.  In all of these sources the dynamical effect of the radio jet on
the line-emitting gas is considerable with major axis velocity gradients of up 
to 1000\kms\ and line widths $\simgt$ 1000\kms.  A further characteristic of 
these radio source---ambient medium interaction objects is the relatively 
close correlation between the radio and line-emission morphologies. 
In \C{34} this is obviously not the case, with the size of the radio source
exceeding the regions of \Oii\ emission by more than a factor of two.

A critical test of a direct interaction between the radio source working
surface and the gas that it excites and compresses are the implied cooling
times.  Assuming an angle between 
the line-of-sight and the radio source axis of 75\degr\ (\cite{Johnson95}), 
the deprojected flow velocity near component $c$ would be
370\kms\ $\times$ cos$^{-1}$(75\degr)~$\simeq$ 1400\kms.
%
%The projected velocity at the position of component $c$ is 367\kms.
%
%The maximum $\Delta$v near this component, however, is $<$400\kms.  
%So, in the interest of obtaining an upper limit to the cooling
%time allowed, we will assume that the higher flow velocity is representative
%of the velocity of the interacting bow shock.
By relating the shock temperature to the temperature of the ambient external
medium and the mach number of their highly supersonic model, \cite{MH92}
deduce an expression that relates the advance speed of the radio source bow
shock ($V_{\rm bow}$) to the temperature of the shock ($T_{\rm sh}$),
\begin{equation}
\label{eqn:shock_temp}
V_{bow} = 0.28~~T_{\rm sh}^{1/2}~~~~\kms.
\end{equation}
Using this relationship we estimate that such a shock will heat the gas in
\C{34} to $T_{\rm sh}\sim~$2.5$\times$10$^7$~K.
In terms of this temperature immediately behind the bow shock and the density 
$n_c$ in the shocked gas clouds, the cooling time is given by
\begin{equation}
\label{eqn:cooling_time}
%\tau_{\rm cool} = \frac{E}{|dE/dt|} = 1.94~\frac{3}{2}~
%              \frac{kT_{\rm sh}}{\Lambda (T) n_c}
%              \simeq 1.27 ~\frac{T_{\rm sh}}{n_c}~~~{\rm years}.
%
\tau_{\rm cool} = \frac{E}{|dE/dt|}
              \simeq 1.27 ~\frac{T_{\rm sh}}{n_c}~~~{\rm years}.
\end{equation}
For the low metallicities expected for the IGM around high redshift radio
galaxies 
%($Z/Z_\odot \simlt$ 10$^{-2}$) 
and shock temperatures
$\simgt$ 10$^7$~K, we have assumed a constant cooling coefficient
$\Lambda (T) \simeq$ 10$^{-23}$\ergcms\ (\cite{Bohringer89}).
Using a canonical ambient medium post-shock density of 
$n_c\sim$~1 -- 10~cm$^{-3}$ along with the derived shock temperature in equation 
\ref{eqn:cooling_time}, we can then compute the time required for the shocked 
plasma to cool to about 2 $\times$ 10$^4$~K.  We find that
$\tau_{\rm cool}\sim$~3 $\times$ 10$^7$ -- 3 $\times$ 10$^6$ years are needed 
before this region can emit in its \OII\ line.  The dynamical age of the 
interaction, however, as derived from the estimated shock velocity and the 
distance between component $c$ and 
%the supposed source of its shock heating,
the radio source hotspot (d~$\simeq$~150\kpc), is on the order of 
1 $\times$ 10$^8$~years.  Thus, with 
$\tau_{\rm dyn}/\tau_{\rm cool}\simgt$~3 -- 30 for the eastern emission-line 
component the gas would have since cooled, and with the short radiative 
lifetime of its current line-emission luminosity (see below), cannot now be 
radiating the energy given to it by the radio source bow shock.

Of course, an alternative to a one-time energy injection could consist of
secondary shocks driven into the cocoon surrounding the supersonic jet itself,
or a more recent input of energy via the lateral expansion of the radio lobes.
Although we cannot entirely discount such models, and it is certainly possible
that some fraction of the excitation of the eastern line-emission region is
due to collisional excitation by the radio lobe, the extremely quiescent velocity
widths and large luminosity in this \Oii\ gas, are difficult to explain solely 
with this interpretation.  
For example, by assuming that the velocity dispersion of the \OII\ gas
is representative of the material that constitutes the bulk of the ionized gas 
mass, the total kinetic energy of this medium will be of the order
\begin{equation}
\label{eqn:kinetic_energy_3c34}
E_k = 1 \times 10^{55}~ M_8~ \sigma^2_{100}~~~~{\rm erg},
%
% prefix coefficient = 9.945 x 10^54
%
\end{equation}
where $M_8$ is the mass of the ionized gas in the eastern region in units of 
10$^8$~\Msun, and $\sigma_{100}$ is the total velocity
dispersion in units of 100\kms.  For the eastern line-emission region we find
$M_{\rm gas}$ = 3.4 $\times$ 10$^8$.  If we further assume that the observed
radial line widths in this region are solely the result of turbulent motions
in an isotropic system, $\sigma$ = $\sqrt 3\, \cdot\,$130\kms.  This means 
that the kinetic energy of the warm ionized gas is of the order of 
2 $\times$ 10$^{56}$~erg.  
Given the total luminosity of this region one derives a radiative lifetime of 
the line-emission gas of
\begin{equation}
\label{eqn:radiative_lifetime_3c34}
\tau = 3 \times 10^4~\frac{E_{k\,56}}{L_{\rm lines\,44}}~\epsilon~~~~{\rm years},
\end{equation}
where $E_{k\,56}$ is the kinetic energy in units of 10$^{56}$ erg,
$L_{\rm lines\,44}$ the luminosity of all the emission-lines in units
of 10$^{44}$\ergs, and $\epsilon$ the efficiency factor in converting
turbulence to line-emission.  So, with conversion efficiencies ranging
from 1 to 100\% the dissipation time of the shock excited gas is only 
140 to 1.4 $\times$ 10$^4$ years.  
%Therefore, such shocks must be
%of sufficiently high frequency to be continuously supplying the line-emitting
%gas with energy, or the medium must be extremely efficient in converting these
%shocks into radiative energy.  

\subsubsection{Photoionization by a Jet-Induced Starburst}

Since most high redshift radio galaxies have blue colours, perhaps indicative
of large scale bursts of star formation, it has been suggested that the
direct interaction between the outflow along the radio source axis and the
IGM is the trigger for this phenomena (\eg\ \cite{McCarBreug87} and
\cite{ChamMil87}).  Although it is not a general phenomena, there is
evidence that jet-induced star formation does indeed occur in low redshift
sources such as Minkowski's object (\cite{BreugFili85}).  Aside from
providing a natural explanation for the radio/optical alignments, this model
also provides a means of delivering ionizing radiation directly to the
line-emitting gas.  Therefore, the large extent of its line-emission region and 
its relatively close alignment to the radio axis makes \C{34} a possible 
candidate for the jet-induced starburst scenario.  

To see whether this model is compatible with our observations,
we will use the spectral evolution models of \cite{Bruzual93} to test
whether the OB stars, produced by the passage of the radio source working
surface, can provide a sufficient number of ionizing photons to {\it in situ}
excite the surrounding line-emitting regions.  For simplicity we have chosen 
an instantaneous burst model with a Salpeter initial mass function (IMF) 
(\cite{Salpeter55}) that is weighted toward the high stellar mass end such 
that 2.5 $\le$ M $\le$ 125~$\Msun$.  Though   
extreme by comparison to other available models,
the choice of an instantaneous starburst is made in order to provide the
greatest amount of UV photons per stellar mass involved.  The same is
true for the choice of IMF and stellar mass range.  Furthermore, the
short duration of the radio source phenomena forces the need to convert the
gas, shocked and compressed by the radio jet or its pressure cocoon, into
stars in the shortest time possible.
A maximum number of ionizing photons in the galaxy rest frame are created
by this model at a time $\tau \sim$ 3.3 $\times$ 10$^{6}$ years after the
initial burst.  By integrating the resultant starburst spectrum blueward 
of the \Lya\ line we find that 
5.1 $\times$ 10$^{47}$ photons~s$^{-1}$~$\Msun^{-1}$ are available to 
photoionize the ambient gaseous medium per solar mass of the starburst.

The \OII\ luminosities observed in the line-emission components $B$, $a$, $b$,
and $c$, the likeliest sites for the location of an underlying starburst, can 
then be divided by this number to determine the minimum total mass of the 
starburst necessary to produce the required ionization.  The computed masses 
range from 7 $\times$ 10$^{6}$ \Msun\ for component $b$, to 
3 $\times$ 10$^{7}$ \Msun\ for source $B$.  Barring 
extinction effects in the environment of the radio galaxy, which would 
undoubtably also affect the observed line-emission, these masses can be
converted to observed flux levels by integrating the starburst spectrum over our
the wavelength range sampled by our continuum filter 
($\lambda_{\circ}$=4033\AA\ (\delam=107\AA)).  The resulting
flux values have been converted to counts in this filter and added to our 
continuum image of \C{34}.  Figure \ref{fig:3c34starburst} shows these
model starbursts to be very obvious at the positions of components $B$ and $c$.
(The more subtle enhancements at the positions of $a$ and $b$ are due to the
weaker line-emission and pre-existing continuum at these positions).
The absence of such obvious starburst continuum signatures in the original 
image (see figure \ref{fig:3c34_lab_comp}b), near sources $B$ and $c$,
strongly excludes this model for the ionization of the line-emission regions 
in \C{34}. 
%
%\placefigure{fig:3c34starburst}
%
\vspace*{-5mm}

\section{Summary and Conclusions}

Our morphological and kinematical study of \C{34} favours a scenario in
which the extended emission-line region is excited by the photoionizing
UV radiation emitted by the central, hidden AGN.  This model goes far in
explaining the overall conical morphology of \C{34}'s warm gas, and the
detailed shapes of its individual \Oii\ knots.  Furthermore, the energy budget
necessary to excite this extended gas remains compatible with the central
luminosity of typical 3C quasars at the same redshift.  To account for the
observed line-emission kinematics, however, requires an 
additional weak interaction between the radio lobes and the ambient medium.
The large, quiescent, blueshifted velocity structure across the entire 
eastern region of \C{34}, as well as the redshifted velocity discontinuity
of the western region, are best explained as gas swept aside by the lateral 
expansion of the radio cocoon. 
This is corroborated by the spatial coincidence between the edges of the
radio lobes and the \Oii\ gas, and by the radio depolarization at positions
where this gas is in direct contact with the radio lobes.  
In this model the alignment of the \Oii\ region with the radio source arises 
from both emission signatures originating from the same central engine, and by 
the fact that the ionizing radiation preferentially escapes along an axis in 
which the ambient medium has been cleared by the radio source.  
The identification of \C{34} with a cone of photoionization may be the first 
detection of such a feature in a high redshift, powerful radio galaxy.
As such, this source can be an important link to the 
low redshift, low luminosity Seyfert galaxies in which such photoionization 
cones have traditionally been observed.

It should be noted that the line-emission morphology and kinematics of \C{34}
are unique in our current \Oii\ sample of 11 radio galaxies.  This means that
we cannot claim photoionization by a hidden, central AGN to be a general 
mechanism for exciting the line-emission regions observed in all powerful 
radio galaxies.  
In fact, at least two further means of ionization are necessary to 
explain the line-emission
morphology and kinematics of all the sources in our sample: shock excitation
of the ambient gaseous medium by an interaction with the radio source
(\cite{MH92}; \cite{HM92}), and strong galaxy--galaxy interactions 
(\cite{Neeser97}).  
This argues against any universal, all-encompassing model
for the alignment effect in powerful radio galaxies;
a fact also supported by the need for both photoionization and 
radio source expansion and backflow to explain most aspects of \C{34}'s 
line-emission regions.

\acknowledgments

This research has made use of the nasa/ipac extragalactic database (ned)
which is operated by the jet propulsion laboratory, caltech, under contract
with the national aeronautics and space administration.  MJN thanks the
anonymous referee for a careful reading of the manuscipt and many helpful
suggestions.

\clearpage

\appendix

\section{Photoionization by a Hidden AGN}

A critical feature of a photoionization model is that the central AGN be able to 
provide a sufficient number of ionizing photons, to satisfy the total luminosity 
observed from the line-emission regions it is supposedly exciting.
The number of ionizing photons emitted by the hidden AGN into a cone of
solid angle, $\Omega_{\rm beam}$, and intercepted by a cloud at a distance
$R$ from the central source is,
\begin{equation}
\label{eqn:NionAGN1}
{\cal N}_{\rm ion}(\nu) = \frac{A}{\Omega_{\rm beam}R^2}~\int_{\nu_c}^{\infty} 
\frac{L_{\rm ion}(\nu)}{h\nu}~d\nu.
\end{equation}
Here, $L_{\rm ion}(\nu)$ is the ionizing luminosity emitted by the AGN, $A$ is 
the effective area of the cloud as seen from the AGN, and the integration 
is over all frequencies above the Lyman-continuum ($\nu_c$).  We assume that 
the ionizing radiation emitted by the central AGN obeys a non-thermal power-law
of the form 
\begin{equation}
\label{eqn:L_power_law}
L_{\rm ion}(\nu) = C\nu^{-\alpha} ~~~~~	(\alpha > 0),
\end{equation}
where $C$ is a constant and $\alpha$ is the spectral index of the far--UV 
energy distribution of the central source.
Equation \ref{eqn:L_power_law} can then be substituted into equation 
\ref{eqn:NionAGN1} and integrated such that,
\begin{equation}
\label{eqn:NionAGN2}
{\cal N}_{\rm ion} = \frac{AC\nu_c^{-\alpha}}{\Omega_{\rm beam}R^2h\alpha}.
\end{equation}
This number of `expected' ionizing photons must then be compared with the
`observed' number of \OII\ photons emitted by the cloud, ${\cal N}_{\rm [OII]}$.
However, to relate the incident radiation
to the observed \OII\ line-emission we must take into account the ratio, $k$,
of the number of ionizing photons required to produce one observed \OII\
photon.  Furthermore, the effectiveness with which a given clump of gas absorbs
incident photons is critically dependent on the covering factor, $f_c$ ---
the fraction of the sky, as seen by the nucleus, covered by optically thick
%(to the Lyman continuum)
clouds.  Combining these two factors we can then compute
the effective number of ionizing photons impinging on the cloud,
in terms of the observed number of \OII\ photons emitted by the cloud,
\begin{equation}
\label{eqn:N_eff_obs}
{\cal N}_{\rm eff} = \frac{k}{f_c}~\frac{L_{\rm [OII]}}{h\nu_{\rm [OII]}}.
\end{equation}
Comparing equations \ref{eqn:N_eff_obs} and \ref{eqn:NionAGN2} and
solving for the constant $C$, which is then introduced into equation 
\ref{eqn:L_power_law} gives the monochromatic luminosity that the hidden AGN
must emit to produce the observed line-luminosity,
\begin{equation}
\label{eqn:NionAGN3}
L_{\rm ion}(\nu) = \frac{k}{f_c}~\frac{L_{\rm [OII]}}{h\nu_{\rm [OII]}}~~
\frac{\Omega_{\rm beam}R^2h\alpha}{A}\biggl(\frac{\nu}{\nu_c}\biggr)^{-\alpha}.
\end{equation}

Most critical for this photoionization model will be those emission-line
components that have the highest surface brightness and lie furthest from 
the central source.  But, before this equation can be applied to these 
sources a number of further assumptions must be made about the variables in
this expression.  For simplicity, we will assume that the projected area of the 
\Oii\ sources (clouds)
is similar to the actual total area facing towards 
the continuum source, with a circular cross-section having a characteristic 
radius ($r$) of one-half the knot's FWHM.  
%
%The solid angle subtended by the 
%ionization cone ($\Omega_{\rm beam}$) will be defined from the position of the 
%central continuum source in our line-free images, and open outward to include 
%all of the extended line-emission observed in our Fabry-Perot frames.  
%
We must also consider what fraction of the ionizing radiation goes into
producing the observed \OII\ flux.  
Assume that the environment of our radio galaxies can be 
approximated by the radiative transfer properties in a giant extragalactic 
HII region, we can use the model results of \cite{Stasinska80}.  Using
parameters typical of the line-emission gas observed in our sources; 
a temperature of $T_{\rm eff}$=4.5 $\times$ 10$^{4}$, 
a metal abundance relative
to solar of $Z/Z_\odot \simlt$0.2, and interpolating between the computed 
densities
10 $<$ n$_e$ $<$ 100 cm$^{-3}$, this model gives an \OII\ to $H\beta$ ratio
of 1.5.  For case B recombination, the ratio of the number of $Ly\alpha$ to 
$H\beta$ photons is 8.5 (\cite{Osterbrock}).  Therefore, we estimate
that the number of photoionizing photons required for every \OII\ photon
produced is $k\sim$6.  Finally, we will assume a spectral index of 
$\alpha$ = 1 (\cite{VillarMartin97}).
%
% alpha=1:  This paper uses the composite spectrum of 16 radio-loud QSO's
%	    spanning z=0.158--3.78 and finds for lambda=1000-5500A that
%	    alpha=0.70 (I find for lambda=1000--4000A that alpha=1.08.
%	    Hewett (\cite{Hewett91}) just assumes alpha=1.00 and gives
%	    no reference for it!
%
%

With these assumptions in place, equation \ref{eqn:NionAGN3}
can be used to compute the {\em required} luminosity
that the ionizing source must have in order to produce the observed \OII\
flux. By integrating this expression with respect
to $\nu$, over any filter of interest that covers a rest frame wavelength range
of $\lambda_1 < \lambda < \lambda_2$, we can 
%extrapolate to longer wavelengths and
determine the broad--band 
luminosities the AGN must produce.  In this way the required luminosity can
be simplified to contain only those parameters specific to an individual source,
\begin{equation}
\label{eqn:NionAGN4}
L_{\rm ion}^\lambda = 24.5~ L_{\rm [OII]}~ \frac{\Omega_{\rm beam} R^2}{f_c A}~
ln\biggr(\frac{\lambda_2}{\lambda_1}\biggl)~~~\ergs.
\end{equation}
This, in turn, can be converted to
expected apparent magnitudes if the ionizing beam from the hidden ionizing
source were directed toward us. The resulting luminosity and magnitude
can then be compared with values from known 3CR quasars at redshifts similar 
to that of \C{34}.  Since the `unifying schemes' assume radio galaxies to 
differ from quasars primarily in terms of orientation, the luminosity that
we compute to be necessary to photoionize the observed line-emission, must 
be less than or equal to those observed for the intrinsically brightest quasars.

%
%	Now for the figure captions:
%

\clearpage
\figcaption[fig1]
{Labelled components of \C{34} as discussed in text.
(a) Continuum subtracted \OII\ image.
(b) Line-free continuum image with a rest-frame wavelength of
$\lambda_{\circ}$=4033\AA\ (\delam=107\AA).
The first contours are at 2.5$\sigma$ above the mean 
background level, while subsequent contours are at levels of 1.5n$\sigma$.
For all images presented, north is at the top and east is to the left, and the 
centroid of continuum component $A$ defines the origin.
\label{fig:3c34_lab_comp}}
\figcaption[fig2]
{A logarithmic grayscale representation of the line-emission
(top panel) and continuum (bottom panel) images of \C{34}.
The 20\cm\ radio contour map of Neff et al. 1995, whose spatial resolution of 
1\fasec4 FWHM beam is comparable to that of our images,
is superimposed on the \Oii\ frame.  The line-emission image consists of the 
sum of 18 Fabry-Perot frames obtained across the line profile.  The continuum 
image has a rest-frame wavelength of $\lambda_{\circ}$=4033\AA\ (\delam=107\AA).
%and defines the origin of the coordinates in both frames.
\label{fig:3c34_line_cont}}
\figcaption[fig3]
{Intensity cross-sections of \C{34}.
The top three panels show the \Oii\ line-emission for the three major regions
of this source; the eastern emission-line region (components $b$ and $c$),
the central section (components $A$ and $B$), and the western region
(component $a$).  The bottom panel shows the
continuum profiles of component $A$
%in the $\lambda_{\circ}$=4033\AA\ (\delam=107\AA)
and a nearby star (dotted line).  The three cross-section cuts, east, center, 
and west, are shown in the inset line-emission contour plot.
All positions are given as straight-line distances from the central continuum
peak.  It should be noted that 
the x-axis distance scale of each plot, with the exception of the central 
continuum and line-emission plots, are different from one another.
Also, the smoothness of these curves is partly due to the rebinning of
the data, such that only every second data point is independent.
\label{fig:3c34_Xsection}}
\figcaption[fig4]
{\Oii\ line profiles at the positions of the major components labelled
in figure \ref{fig:3c34_lab_comp}, and at the diffuse emission between
components $b$ and $c$ ($b$-$c$).  Each profile
is an average over 4$\times$4 pixels in units of counts per pixel above the
continuum.
\label{fig:3c34profiles}}
\figcaption[fig5]
{A colour representation of the radial velocities 
(top panel) and the deconvolved velocity widths (bottom panel) of the \OII\ 
line-emission in \C{34}.  The axes are labelled in arc seconds from the origin 
defined by the optical continuum emission peak $A$ (shown with a cross).  
At the bottom of each 
panel a gradient scale is given matching the diagram's colours to velocities in
\kms.  On the velocity map we also show the axes used for the cross-sections
given in figure \ref{fig:3c34velfwhmpro}.
The lengths of the cuts are given in figure \ref{fig:3c34_Xsection}.
\label{fig:3c34velwidt}}
\figcaption[fig6]
{The velocity profiles (top panels) and the deconvolved 
velocity widths (bottom panels) along three major axis cuts of \C{34}.  The 
three panels for each of velocity and FWHM are divided into the eastern, central,
and western regions defined in figure \ref{fig:3c34_Xsection}.
The position and velocity origin has been taken to be at the radio galaxy 
center (component $A$).
\label{fig:3c34velfwhmpro}}
\figcaption[fig7]
{Our proposed photoionization cone superimposed on the grayscale \Oii\ 
image of \C{34}.  The dashed line indicates the radio source axis, as measured 
from the outermost hotspots of the 20\cm\ map shown in figure 
\ref{fig:3c34_line_cont}.
The apex of the cone is located
at the position of the central continuum source $A$; the location of the
hidden AGN.  An opening angle of
60\degr\ is the minimum required to photoionize the observed line-emission.
\label{fig:3c34cone}}
\figcaption[fig8]
{A schematic view perpendicular to the line-of-sight to
\C{34}, looking {\em into} the eastern radio lobe of this source.  In our model,
the gas of the eastern line-emission region has been swept up by the southern
end of the expanding radio lobe.  The greater column depth along our 
line-of-sight to the bottom of the line-emission region results in a higher
\Oii\ surface brightness on the outer side of this feature.
\label{fig:3c34_edge_bright}}
\figcaption[fig9]
{An {\bf artificial} continuum image of \C{34} 
%(rest-frame $\lambda_{\circ}$=4033\AA) 
as it would appear if each of the line-emission components were the result of 
an underlying starburst.  Each starburst has been marked with the label 
corresponding to its associated line-emission component.  
With the exception of components $a$ and $b$, in which continuum
emission exists in the original image, components $B$ and $c$ are prominent 
and located in regions previously devoid of comparable continuum emission.
Each starburst was given the same PSF as its associated line-emission
component.  This contour map should be compared with the observed continuum 
image shown in figure \ref{fig:3c34_lab_comp}b.
\label{fig:3c34starburst}}

\clearpage 

\begin{table*}[t]
\begin{center}
\caption{Properties of \C{34} Continuum and \OII\ Components}
\vspace*{0.5cm}
\begin{tabular}{c c c r c c c c c}
%\hline\hline
\tableline\tableline
\noalign{\smallskip}
\multicolumn{1}{c}{comp.}
&\multicolumn{2}{c}{rel. pos.}
&\multicolumn{1}{c}{r$^a$}
&\multicolumn{1}{c}{redshift}
&\multicolumn{1}{c}{m$_{\rm 4033}$}
&\multicolumn{1}{c}{M$_{\rm 4033}$}
&\multicolumn{1}{c}{f(\OII)$^b$}
&\multicolumn{1}{c}{FWHM$^c$} \\
&\multicolumn{1}{c}{$\Delta\alpha$}
&\multicolumn{1}{c}{$\Delta\delta$}
&\multicolumn{1}{c}{kpc}
&
&
&
&\multicolumn{1}{c}{(R$_{aperture}$)}
&\multicolumn{1}{c}{\AA} \\
[1mm]                                                 \noalign{\hrule} \\
{\em All} & ---     & ---      &--- &0.689             &---   &---   &223 (9.6) &38.7 \\
$A$  & 0\farcs0 & 0\farcs0 &0.0 &0.6900$\pm$0.0007 &20.68 &-22.7 &62.3 (1.8) &17.6 \\
$B$  & 1\farcs4 & 1\farcs3 &15.3 &0.6899$\pm$0.0007 &---  &---   &45.3 (1.6) &18.7 \\
$C$  & 0\farcs5 & 4\farcs3 &34.5 & ---             &21.67 &-21.7 &---  &--- \\
$a$  &-4\farcs3 & 0\farcs1 &34.1 &0.693$\pm$0.001 &---  &---   &18.6 (1.4) &19.5 \\
$b$  & 3\farcs6 &-1\farcs1 &30.1 &0.687$\pm$0.001 &---  &---   &12.3 (1.4) &14.4 \\
$c$  & 7\farcs1 &-3\farcs1 &61.4 &0.6881$\pm$0.0008 &---  &---   &29.4 (1.9) &12.6 \\
\noalign{\smallskip}
\hline
\end{tabular} 
\end{center}
\noindent{ }

\footnotesize

\hspace*{0.7cm} $^a$ H$_{\circ}$=50\kmsMpc\, q$_{\circ}$=0.5

\hspace*{0.7cm} $^b$ flux in units of 10$^{-16}$\ergscm (radius of aperture in arcsec)

\hspace*{0.7cm} $^c$ observed FWHM corrected for instrumental broadening
\label{tab:3c34_components}
\end{table*}

\normalsize

%\scriptsize{
\begin{table*}[b]
\begin{center}
\caption{\label{tab:3c34photo_model} Results of Photoionization Calculations
for \C{34}}
\vspace*{0.5cm}
\begin{tabular}{|c|c|c|c|c|c|}
\hline
\multicolumn{1}{|c|}{Source}
&\multicolumn{1}{|c|}{Dist from}
&\multicolumn{1}{|c|}{Cone Opening}
&\multicolumn{1}{|c|}{Cloud}
&\multicolumn{1}{|c|}{L$_B$ AGN (\ergs)}
&\multicolumn{1}{|c|}{m$_v$ AGN} \\
&\multicolumn{1}{|c|}{AGN ({\tt "})}
&\multicolumn{1}{|c|}{Angle}
&\multicolumn{1}{|c|}{Radius ({\tt "})}
&& \\ \hline\hline
    &    &60\degr &1.02 &1.4 $\times$ 10$^{44}$ &19.5 \\ \cline{4-6}
$B$ &1.9 &	&1.60 &9.3 $\times$ 10$^{43}$ &20.0 \\ \cline{3-6}
    &    &88\degr &1.02 &2.8 $\times$ 10$^{44}$ &18.8 \\ \cline{4-6}
    &    &	&1.60 &1.8 $\times$ 10$^{44}$ &19.3 \\ \cline{1-6}
    &    &60\degr &1.20 &1.1 $\times$ 10$^{45}$ &17.3 \\ \cline{4-6}
$c$ &8.2 &	&1.90 &8.0 $\times$ 10$^{44}$ &17.6 \\ \cline{3-6}
    &    &88\degr &1.20 &2.2 $\times$ 10$^{45}$ &16.5 \\ \cline{4-6}
    &    &	&1.90 &1.5 $\times$ 10$^{45}$ &16.9 \\ \cline{1-6}
\noalign{\smallskip}
\end{tabular}
\end{center}
\end{table*}

\clearpage

%
%	Now the figures themselves:
%
%

%\setcounter{page}{40}

\begin{figure*}[h]
\centerline{
\psfig{figure=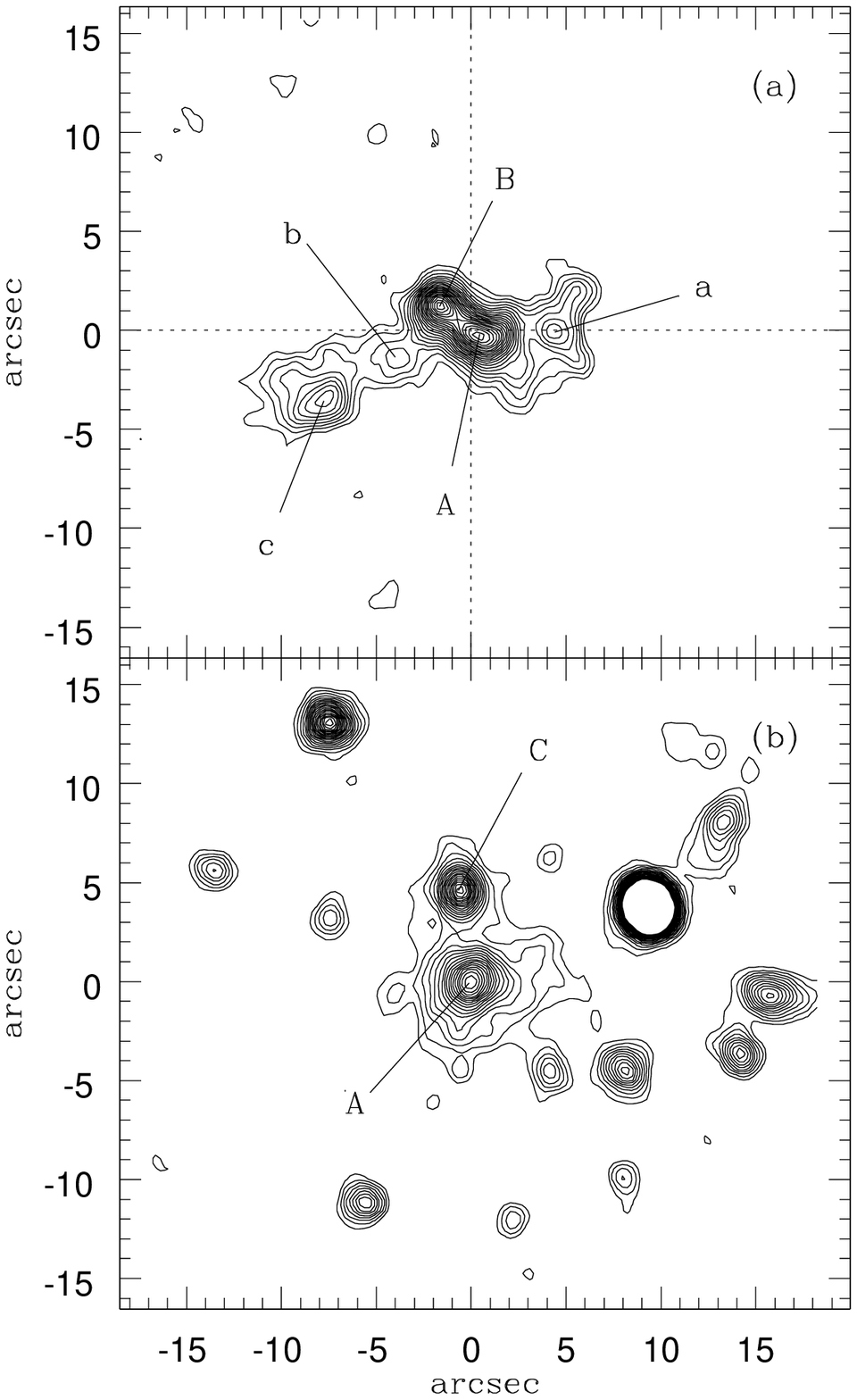,width=22cm}}
\begin{flushright}
{\bf Figure 1:}\ \ Neeser et al.
\end{flushright}
\end{figure*}

\clearpage

\begin{figure*}[h]
\centerline{
\psfig{figure=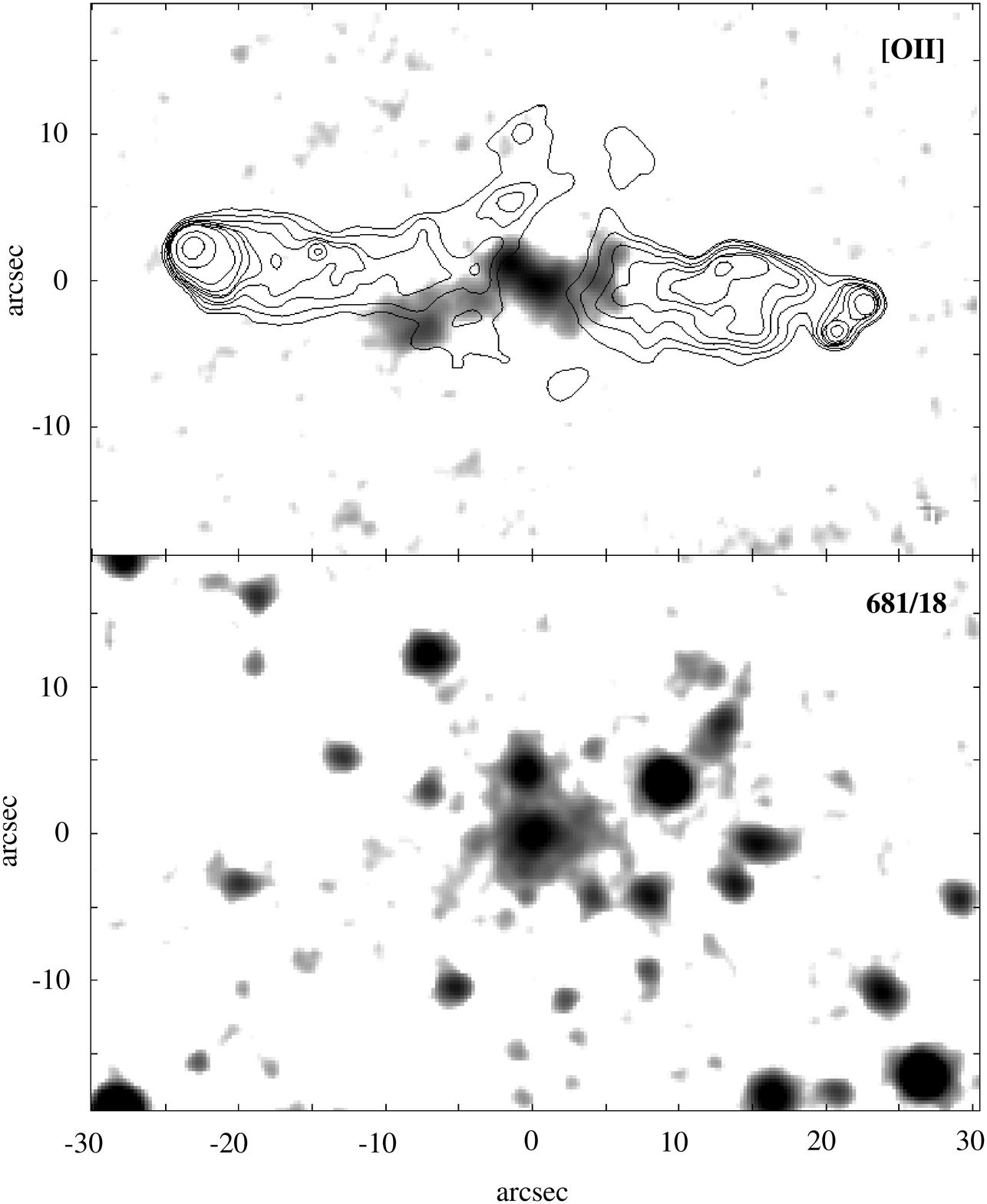,width=18.5cm}}
\vspace*{0.6cm}
\begin{flushright}
{\bf Figure 2:}\ \ Neeser et al.
\end{flushright}
\end{figure*}

\clearpage

\begin{figure*}[tp]
\centerline{
\psfig{figure=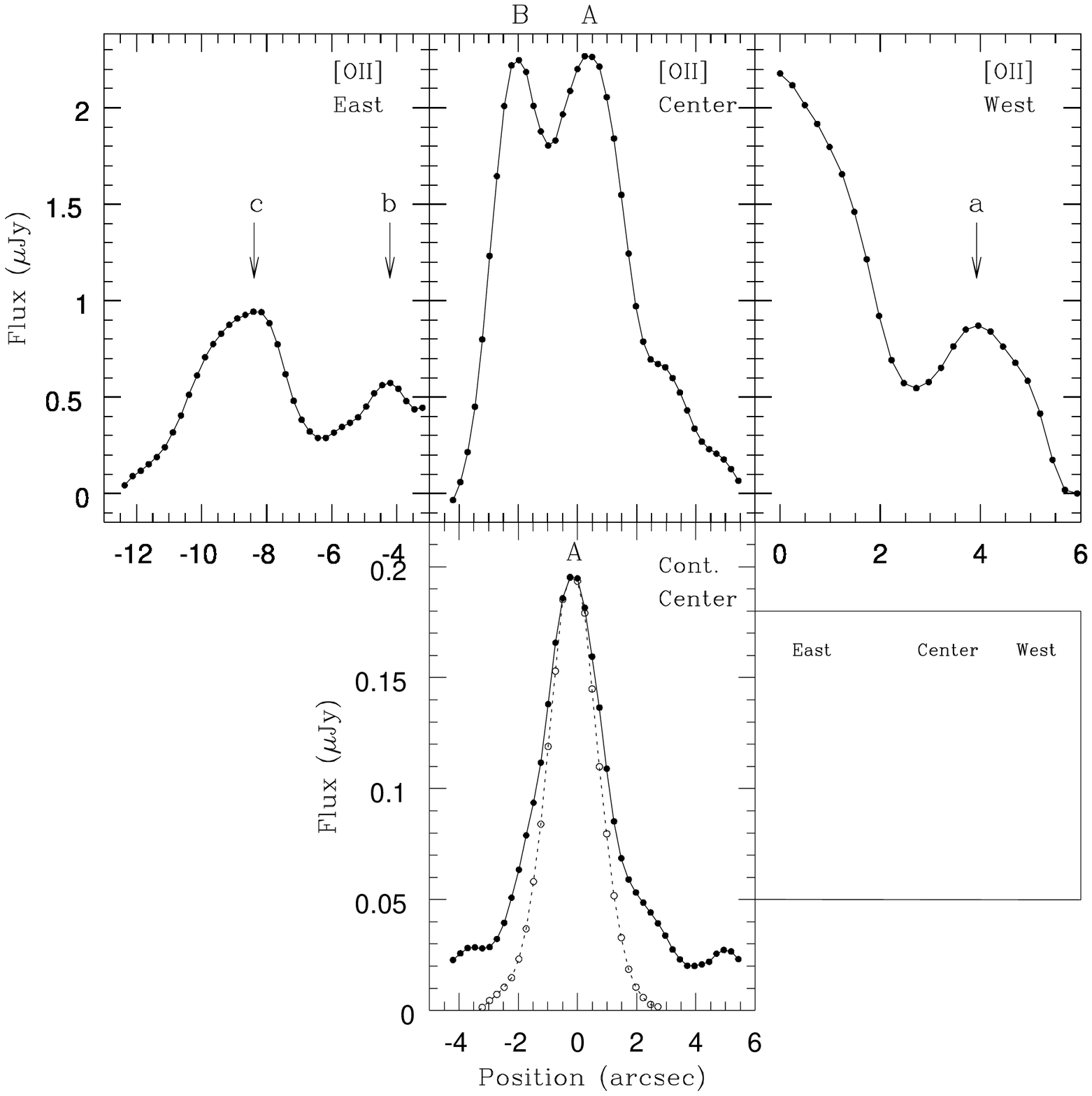,width=19.0cm,height=19.0cm}}
\vspace*{0.6cm}
\begin{flushright}
{\bf Figure 3:}\ \ Neeser et al.
\end{flushright}
\end{figure*}

\begin{figure*}
\includegraphics{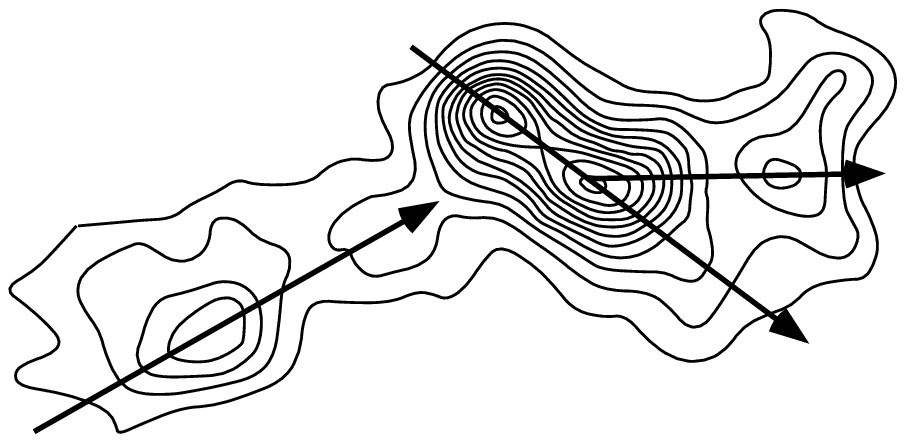}
\end{figure*}

\clearpage

\begin{figure*}[h]
\centerline{
\psfig{figure=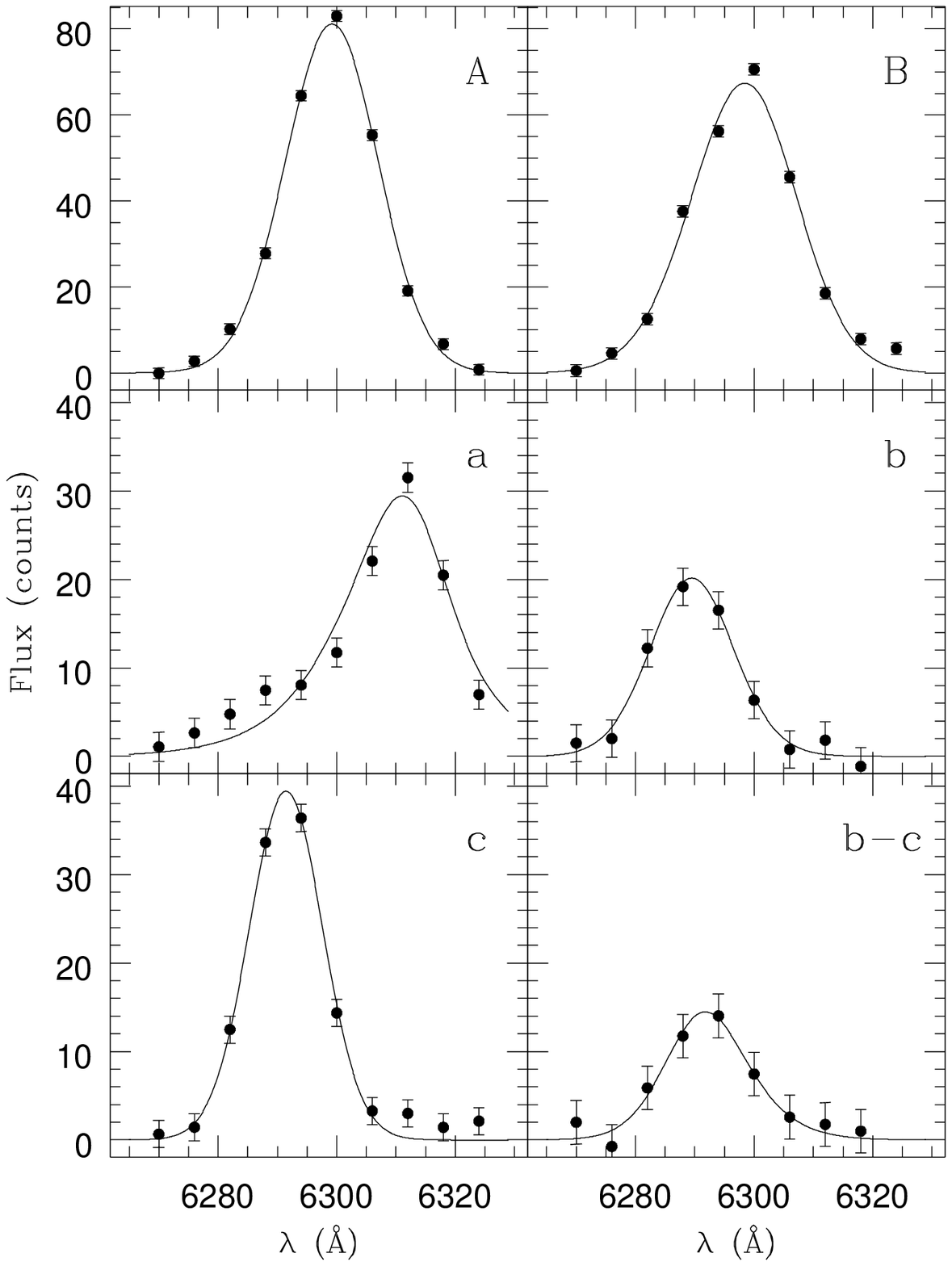,height=20.0cm}}
\vspace*{2.0cm}
\begin{flushright}
{\bf Figure 4:}\ \ Neeser et al.
\end{flushright}
\end{figure*}

\clearpage

\begin{figure*}[h]
\centerline{
\psfig{figure=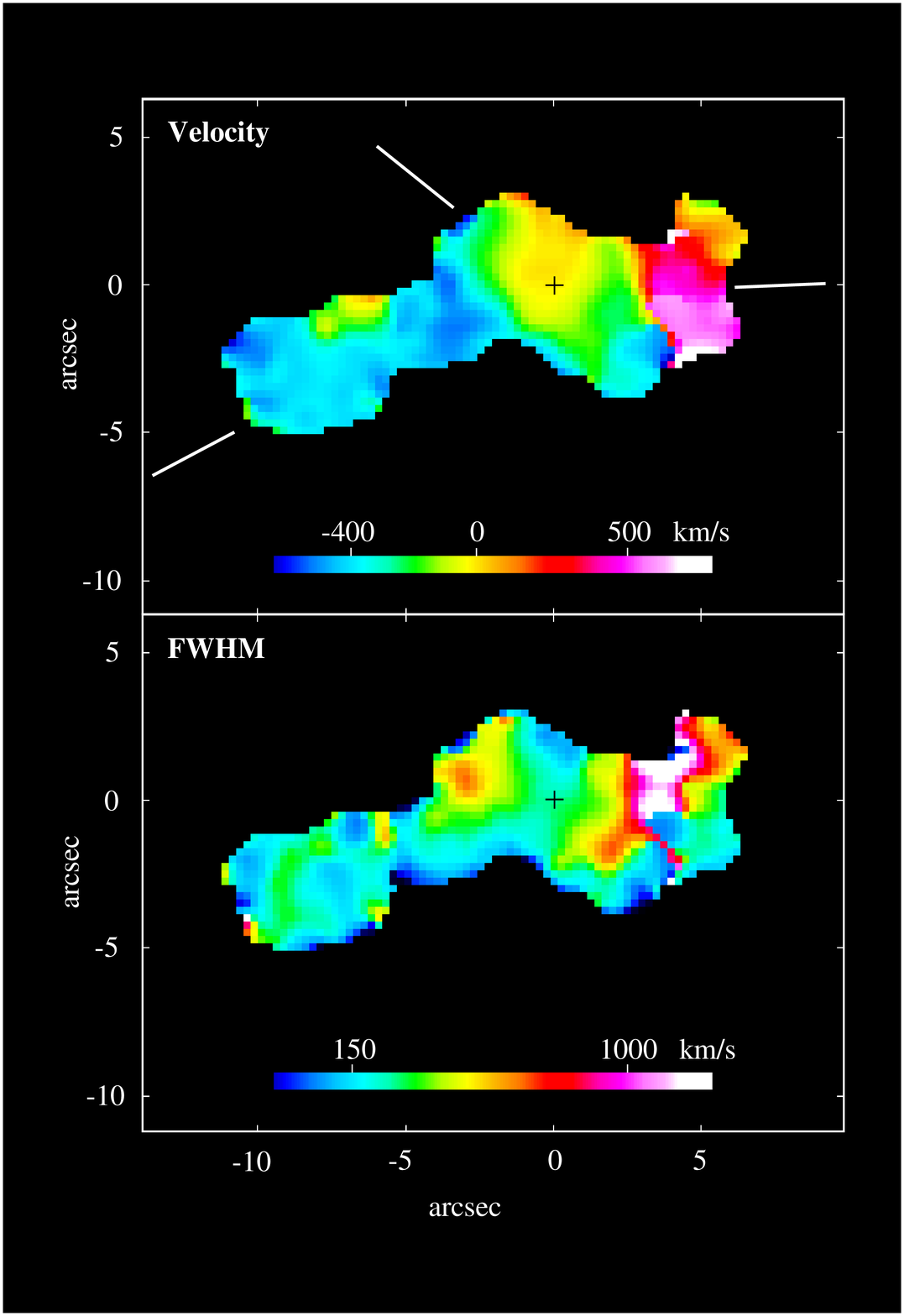,width=15.2cm}}
%\vspace*{0.3cm}
\begin{flushright}
{\bf Figure 5:}\ \ Neeser et al.
\end{flushright}
\end{figure*}

\clearpage

\begin{figure*}[h]
\centerline{
\psfig{figure=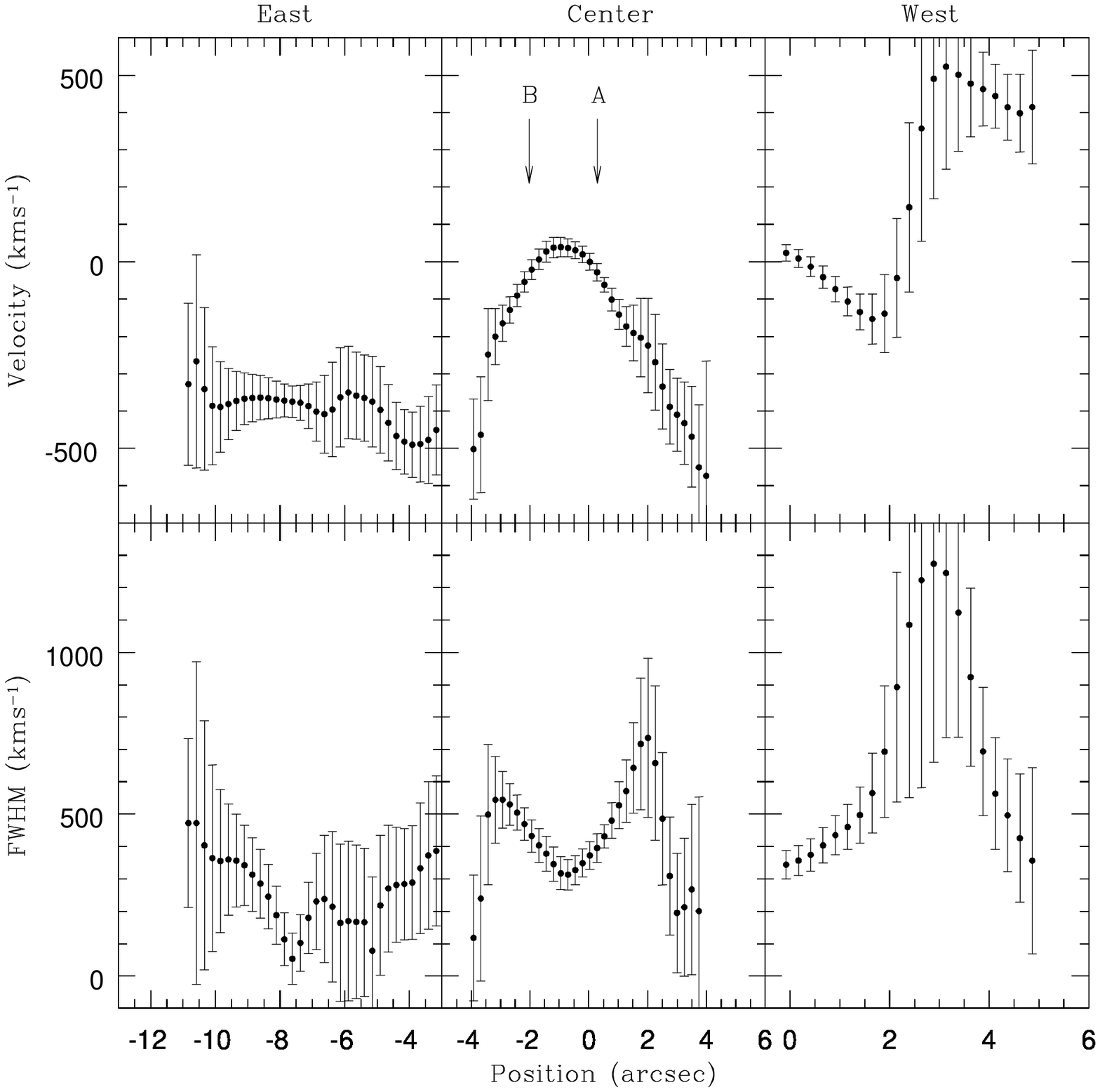,width=17.9cm,height=18.4cm}}
\vspace*{3.3cm}
\begin{flushright}
{\bf Figure 6:}\ \ Neeser et al.
\end{flushright}
\end{figure*}

\clearpage

\begin{figure*}[h]
\centerline{
\psfig{figure=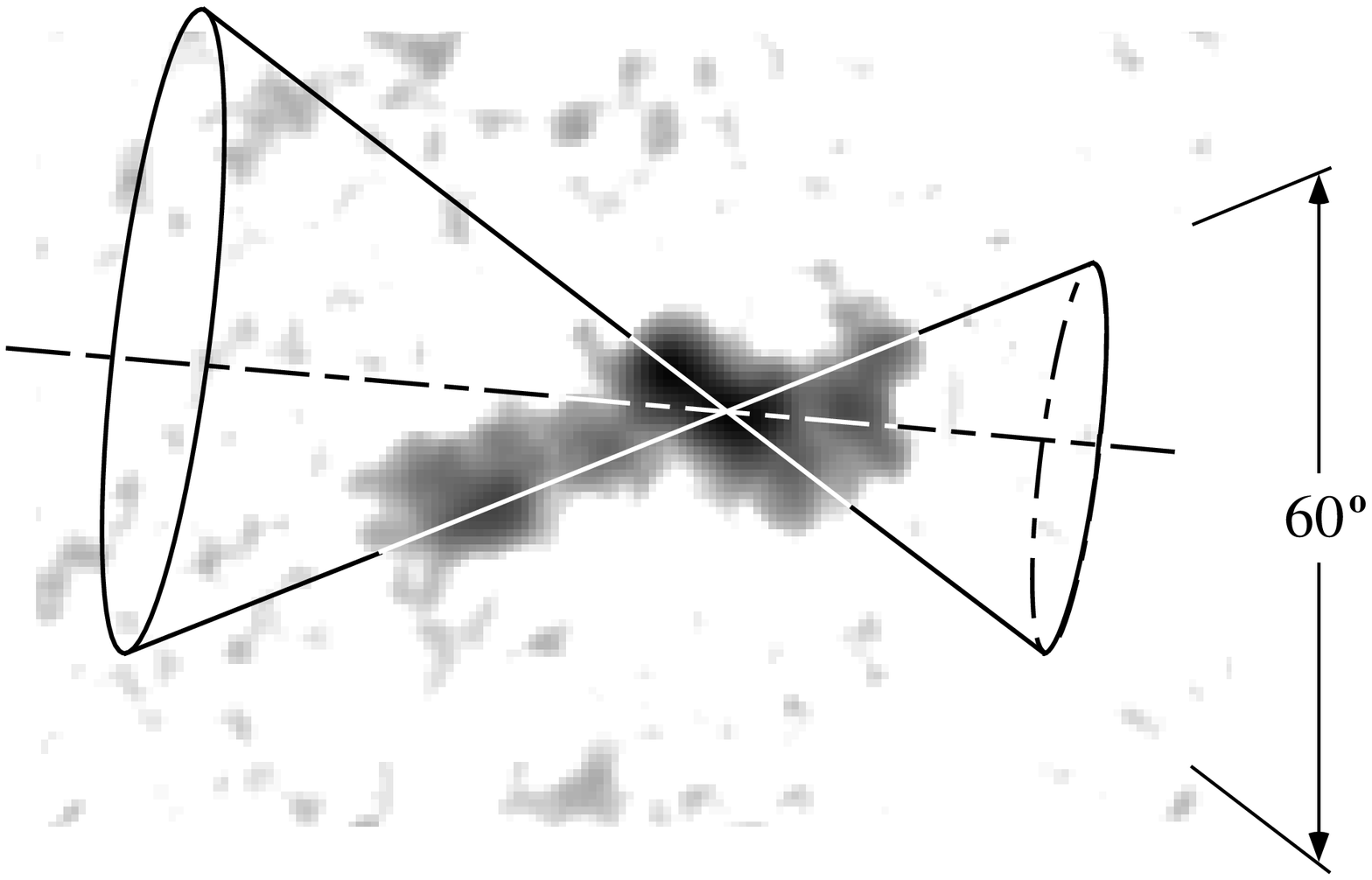,width=20.0cm}}
\vspace*{6.8cm}
\begin{flushright}
{\bf Figure 7:}\ \ Neeser et al.
\end{flushright}
\end{figure*}

\clearpage

\begin{figure*}[h]
\centerline{
\psfig{figure=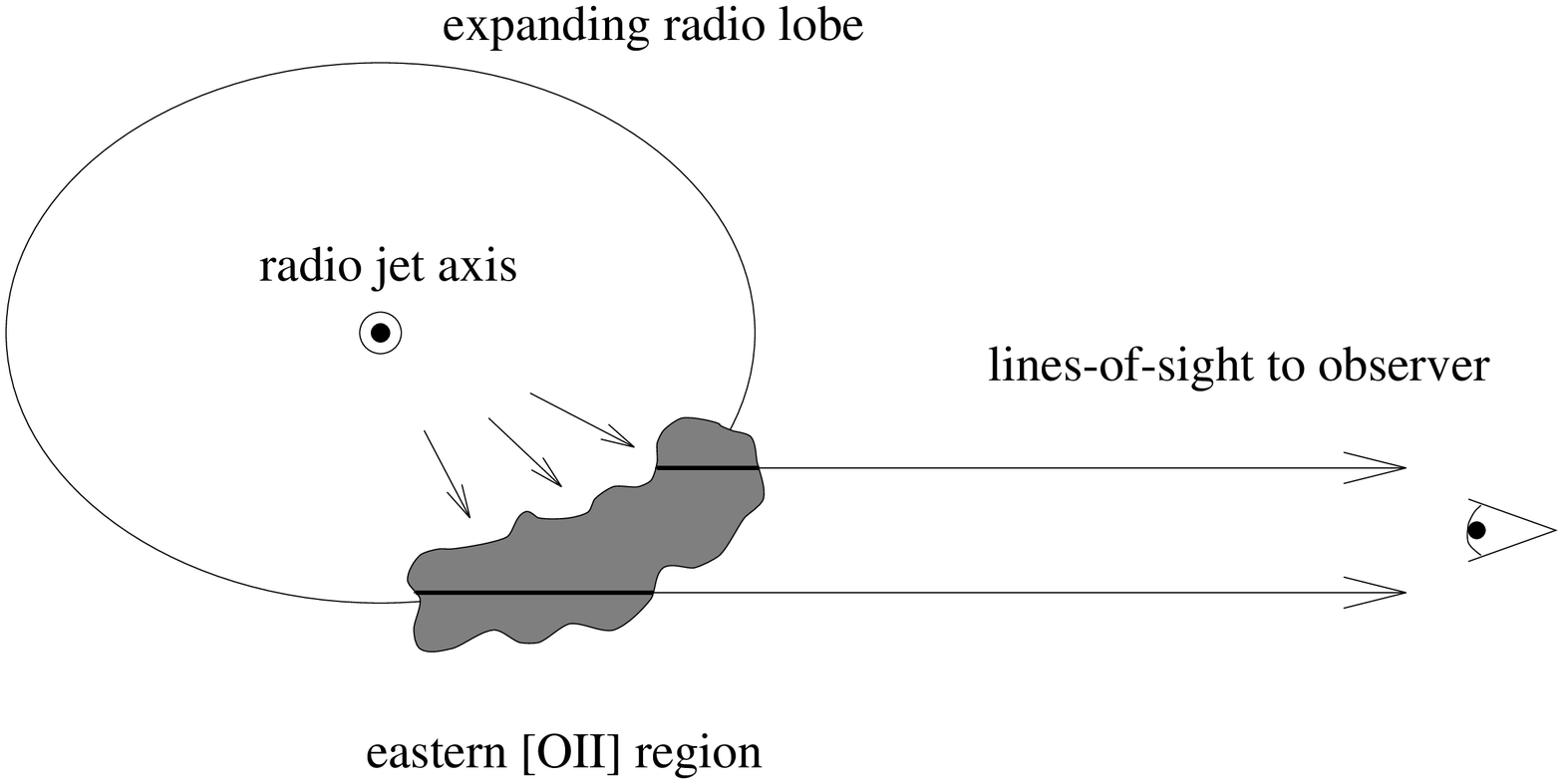,width=16.0cm}}
\vspace*{12.5cm}
\begin{flushright}
{\bf Figure 8:}\ \ Neeser et al.
\end{flushright}
\end{figure*}

\clearpage

\begin{figure*}[h]
\centerline{
\psfig{figure=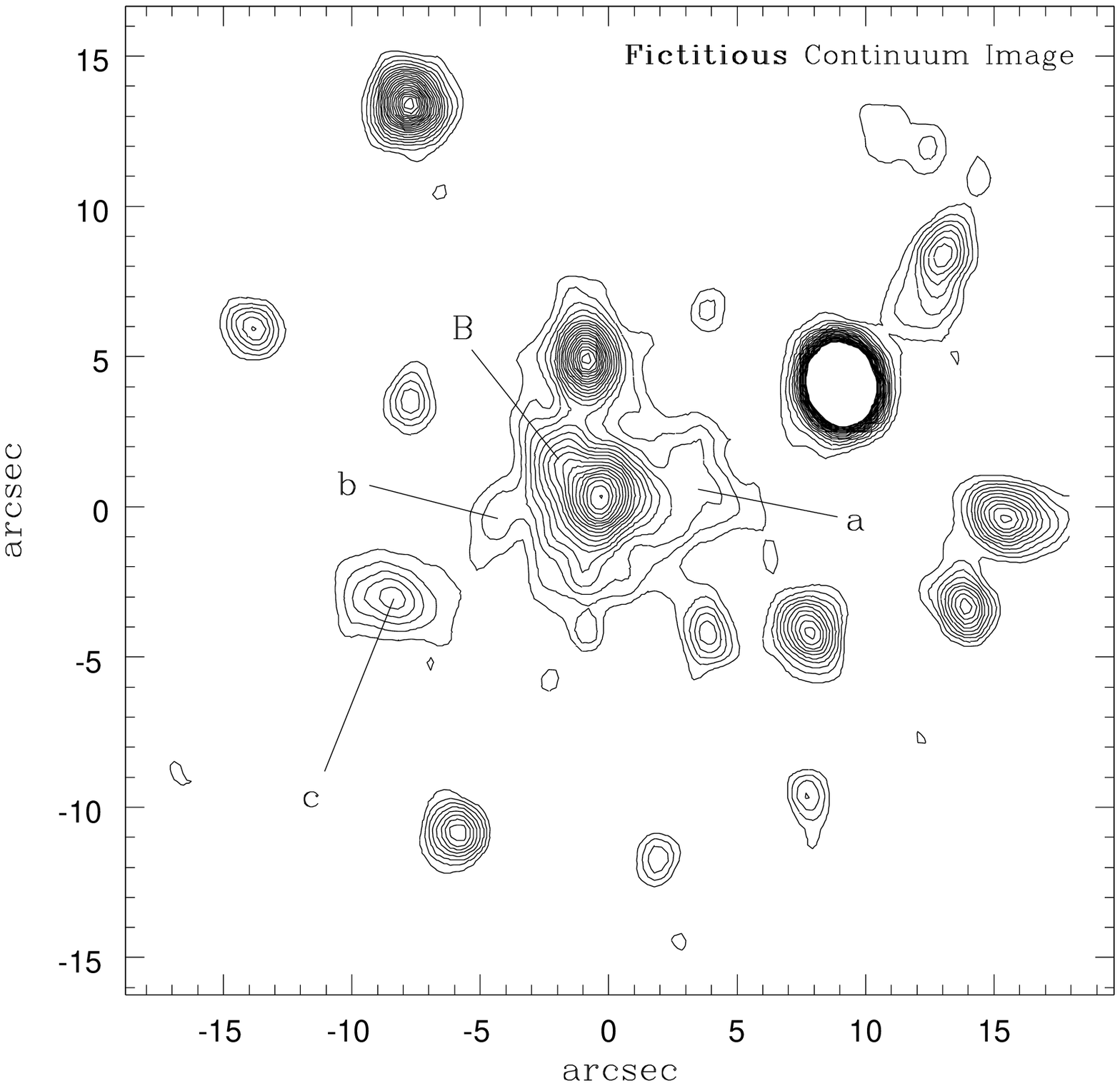,height=15.5cm,width=17.5180cm}}
\vspace*{6.2cm}
\begin{flushright}
{\bf Figure 9:}\ \ Neeser et al.
\end{flushright}
\end{figure*}

\end{document}